\documentclass[aps,prc,nofootinbib,twocolumn]{revtex4-1}

\usepackage{graphicx}
\usepackage{epsfig}
\usepackage{amsmath}
\newcommand{\midtilde}{\raisebox{-0.25\baselineskip}{\textasciitilde}}

\begin{document}

\title{Nucleon self-energies for supernova equations of state}

\author{Matthias Hempel}
\email{matthias.hempel@unibas.ch}
\affiliation{Department of Physics, University of Basel, Klingelbergstrasse 82, 4056 Basel, Switzerland}

\date{\today}

\begin{abstract}
Nucleon self-energies and interaction potentials in supernova (SN) matter, which are known to have an important effect on nucleosynthesis conditions in SN ejecta are investigated. Corresponding weak charged-current interaction rates with unbound nucleons that are consistent with existing SN equations of state (EOSs) are specified. The nucleon self-energies are made available online as electronic tables. The discussion is mostly restricted to relativistic mean-field models. 

In the first part of the article, the generic properties of this class of models at finite temperature and asymmetry are studied. It is found that the quadratic expansion of the EOS in terms of asymmetry works reasonably well at finite temperatures and deviations originate mostly from the kinetic part. The interaction part of the symmetry energy is found to be almost temperature independent. At low densities, the account of realistic nucleon masses requires the introduction of a linear term in the expansion. Finally, it is shown that the important neutron-to-proton potential difference is given approximately by the asymmetry of the system and the interaction part of the zero-temperature symmetry energy. The results of different interactions are then compared with constraints from nuclear experiments and thereby the possible range of the potential difference is limited.

In the second part, for a certain class of SN EOS models, the formation of nuclei is considered. Only moderate modifications are found for the self-energies of unbound nucleons that enter the weak charged-current interaction rates. This is because in the present approach the binding energies of bound states do not contribute to the single-particle energies of unbound nucleons.
\end{abstract}

\maketitle
\section{Introduction}
Recently it was shown that nucleon interaction potentials modify the evolution of neutrino spectra in core-collapse supernovae (CCSNe) \cite{martinez12,roberts12a,roberts12} and that they influence the deleptonization of newly born proto-neutron stars (PNSs). The effect of the potentials is of particular relevance for the so-called neutrino-driven wind (NDW). The NDW represents the emission of a low-density, high-entropy baryonic gas from the surface of the PNS. It is driven by energy deposition of neutrinos emitted from deeper layers and sets in after the launch of the supernova (SN) explosion and remains active in the first seconds up to minutes. 

The NDW is of great importance for nucleosynthesis of heavy elements, as it has been considered as one of the most promising sites for the so-called r-process (see, for example, the review in Ref.~\cite{arcones2013}). However, sophisticated long-term simulations of CCSNe \cite{fischer10a,huedepohl10} have shown that the matter emitted in the NDW is generally proton rich, allowing only for the so-called $\nu$p-process, which is not able to produce the most heavy nuclei \cite{froehlich06a,froehlich06b,arcones07,roberts10,arcones2013}. In Refs.~\cite{martinez12,roberts12} it was realized that these long-term simulations of the PNS deleptonization phase and the NDW neglected the effect of nuclear interactions in the weak interaction rates with unbound nucleons. Implicitly they were assuming a noninteracting gas of (unbound) nucleons. This represents a crucial simplification and is inconsistent with the nuclear equation of state (EOS) used in the same simulations for the thermodynamic quantities. 

For the early phases of a CCSN, like the collapse of the progenitor star, the subsequent accretion phase, and the onset of the explosion, the neutrino spheres are at such low densities that the neglect of the nucleon interactions in the weak rates with unbound nucleons is justified. However, for the later phases of the evolution, when the neutrino spheres move to high densities, this is not the case anymore. The neutrino spectra are modified by the nucleon interactions. An important quantity to characterize this effect is the difference of the (nonrelativistic) mean-field potentials of neutrons and protons, 
\begin{equation}
\Delta U =U_n -U_p \; . \label{deltauu}
\end{equation}
If one thinks of a single reaction, the potential difference increases the energy of an emitted antineutrino and decreases the one of an emitted neutrino compared to a noninteracting gas.

The recent simulations of Refs.~\cite{martinez12,roberts12a,roberts12} have indeed shown that, taking the nucleon interactions consistently into account, this leads to an increase in the difference of the mean energies of neutrinos and antineutrinos. 
This energy difference is a crucial quantity for nucleosynthesis, as only a difference larger than $4 Q$, with $Q=m_n-m_p\simeq1.29$ MeV, would lead to neutron-rich conditions \cite{qian96}. By taking the potential difference of unbound nucleons into account, in Refs.~\cite{martinez12,roberts12} slightly neutron-rich conditions were obtained in the NDW.

Obviously, these results depend on the nuclear interactions being used. The two aforementioned simulations just started to explore the effect of different interactions. In Ref.~\cite{martinez12}, two different relativistic mean-field (RMF) models were used, GM3 \cite{glendenning91} and the more recent IUFSU \cite{fattoyev2010}. In these simulations, the formation of nuclei was not taken into account in the EOS; i.e., only nucleons were considered as degrees of freedom. In addition, the wind was not part of the hydrodynamic simulation. The asymptotic electron fraction in the wind was estimated employing the results of Ref.~\cite{qian96}. In the simulation of Ref.~\cite{martinez12} the wind is part of the computational domain, and the EOS of Shen \textit{et al.}~\cite{shen98}, which is based on the RMF interactions TM1 \cite{Suga94}, includes $\alpha$ particles and a representative heavy nucleus. In both works \cite{martinez12,roberts12} it was pointed out that $\Delta U$ is related to a basic property of the nuclear EOS, the symmetry energy, but no further details were given.

It is one of the main motivations of the present article to investigate the connection between the potential difference $\Delta U$ and the symmetry energy. It will be shown that $\Delta U$ is given by the potential or interaction part of the symmetry energy. Next, different predictions for $\Delta U$ are compared, obtained from all of the RMF interactions which are currently available for use in CCSN simulations. The results of these EOSs are also compared with existing theoretical and experimental constraints, to limit the possible range of values $\Delta U$ could have. 

The second part of the article deals with effects of nuclei on unbound nucleons. The existing studies about the impact of the nucleon potentials on nucleosynthesis conditions in the wind were mostly concentrating on the nucleon component of the emissivity and/or absorptivity. However, in SN matter, one has not only a uniform gas of interacting nucleons, but there is also an important contribution from nuclei. During the collapse, and in the matter which is subsequently accreted onto the shock, heavy nuclei dominate the composition. Also in the matter behind the shock and in the envelope of the newly born PNS, nuclei are present with significant abundances. These are mostly light nuclei like deuterons, tritons, or $\alpha$ particles \cite{oconnor07,arcones08,sumiyoshi08,hempel11,fischer14}. Their effect in the neutrino transport is very interesting as they could potentially modify the neutrino spectra \cite{arcones08}. So far there are only few exploratory studies, e.g., the one of Ref.~\cite{furusawa2013}, that directly incorporate selected neutrino interactions with light nuclei in CCSN simulations.

This article prepares further steps in this direction, by investigating how the appearance of nuclei modifies single-particle properties of unbound nucleons that are relevant for the neutrino interaction rates with unbound nucleons. However, the important aspect of the neutrino interactions with nucleons bound in nuclei is not addressed here. Nevertheless, at least a description of the neutrino reactions with unbound nucleons that is consistent with the underlying EOS for models which are already used in numerical astrophysical investigations is provided. 

The structure of this article is as follows. Section \ref{sec_nucleonic} is restricted to the discussion of uniform nucleonic matter. A review of the formal structure of typical RMF models is given and their temperature and asymmetry dependence is investigated. It is shown that the nucleon potential difference is approximately proportional to the asymmetry of the system and the interaction part of the zero-temperature symmetry energy. The theoretical predictions are also compared with experimental constraints. In Sec.~\ref{sec_snmatter}, the formation of nuclei is considered, as well as what effect they have on the single-particle properties of unbound nucleons and their neutrino interactions rates. Different definitions of the nucleon potential difference are compared and different contributions to the nucleon potential difference are identified. Section \ref{sec_summary} gives a summary, and conclusions are drawn. The Appendix explains the structure of tables that are available online that provide complementary information to current SN EOS tables. They list the self-energies and other single-particle properties of unbound nucleons needed to calculate neutrino interaction rates consistent with the EOS.

\section{Nucleonic matter}
\label{sec_nucleonic}
This section deals with nucleonic matter, i.e., bulk uniform nuclear matter consisting of only neutrons and protons. To derive the connection between $\Delta U$ and the interaction part of the symmetry energy, it is first necessary to summarize some basic and generic properties of RMF models at finite temperature and asymmetry.

\subsection{Relativistic mean-field EOS}
\label{sec_rmf}
Similarly to the potential difference, the neutron-proton mass splitting is important for the NDW. Therefore, it is advantageous to include the mass splitting not only in the neutrino interactions, but also consistently in the EOS. From the RMF models which are considered here, only SFHo and SFHx \cite{steiner13} and DD2 \cite{typel09} are based on real nucleon masses. All other models (TM1 \cite{Suga94}, TMA \cite{toki95}, NL3 \cite{Lala97}, FSUgold \cite{todd2005}, and IUFSU \cite{fattoyev2010}) assume an average nucleon mass with a value in the range from 938 to 939 MeV. In principle, a change of the nucleon masses corresponds to a change of the parameters of the interactions and thus would require a refitting of the model. However, the change of nuclear-matter properties induced by the change of the nucleon masses is small. Therefore, in the present investigation the neutron mass is simply replaced with $m_n = 939.565346$~MeV and the proton mass with $m_p = 938.272013$~MeV \cite{mohr08}, without any refitting. 

In the following, a generic RMF model with momentum-independent interactions and without a scalar isovector interaction is considered. The chosen formalism uses only the scalar and vector self-energies as degrees of freedom, instead of working with the expectation values of the fields. This has the advantage that the description is more independent from the particular Lagrangian used. It is applicable to both conventional meson-exchange-based RMF models with fixed couplings (and possibly nonlinear terms) but also for models with density-dependent couplings. 

In the mean-field picture, nucleons obey Fermi-Dirac statistics and the pressure can be split into a kinetic and an interaction part, $P^{\rm kin}$ and $P^{\rm int}$:
\begin{eqnarray}
 P=P^{\rm kin}+P^{\rm int} +P^{\rm R}\; . \label{psum}
\end{eqnarray}
In addition, for density-dependent models (such as DD2), there is a pressure contribution from rearrangement terms $P^{\rm R}$, to maintain thermodynamic consistency. It contains the terms with derivatives of the couplings with respect to density. Even though $P^{\rm R}$ can also be seen as an interaction term, for the purposes followed here, it is advantageous to distinguish the two contributions $P^{\rm int}$ and $P^{\rm R}$. For models with constant couplings, one has $P^{\rm R}\equiv 0$. In this case also all other quantities with sub- or superscript ``R'' appearing in the following discussion are identical to zero.

The kinetic pressure is given by
\begin{eqnarray}
 P^{\rm kin}&=&\sum_i \frac1{3\pi^2} \int_0^\infty dk \frac{k^4}{E_i^{\rm kin}} (f_i+f_{\bar i}) \; , 
\end{eqnarray}
where $i=n,p$ denotes neutrons and protons, which are the only baryonic degrees of freedom considered in the present section. The distribution functions $f_i$ of the nucleons are
\begin{equation}
 f_i = \frac1{1+\exp[(E_i -\mu_i)/T]} \; . \label{fi}
\end{equation}
For antineutrons and antiprotons one has
\begin{equation}
 f_{\bar i} = \frac1{1+\exp[(E_{\bar i} +\mu_i)/T]} \; , \label{fbari}
\end{equation}
where $\mu_i$ is the corresponding relativistic chemical potential with rest mass included. $E_i$ ($E_{\bar i}$) is the single-particle energy of nucleons (antinucleons). These are given by the momentum $k$, the effective Dirac mass $m_i^*$, and a vector potential generated by the fields, respectively the total RMF vector self-energy of the nucleon $\Sigma_{VR}^{i}$,
\begin{eqnarray}
 E_i &=& E_i^{\rm kin}+\Sigma_{VR}^{i}\; , \label{ei} \\
 E_{\bar i} &=&  E_i^{\rm kin}-\Sigma_{VR}^{i} \; ,\\
E_i^{\rm kin}&=&  \sqrt{k^2+{m_i^*}^2} \; ,
\end{eqnarray}
whereas
\begin{eqnarray}
 m^*_i&=&m_i +\Sigma_S \; ,
\end{eqnarray}
with the nucleon scalar self-energy $\Sigma_S$ and the nucleon vacuum masses $m_i$, for which the experimentally measured values \cite{mohr08} are used, as mentioned above. 
$\Sigma_S$ is assumed to be equal for protons and neutrons. This means scalar isovector interactions are not considered, corresponding to the $\delta$ meson in interaction models that are based on meson exchange. The total nucleon vector self-energy can be separated into a ``bare'' part and one from the rearrangement
\begin{eqnarray}
\Sigma_{VR}^{i}=\Sigma_V^i + \Sigma_R \; . \label{sigmavr}
\end{eqnarray}
$\Sigma_V^i$ is the more important quantity for the present study because $\Sigma_R$ is isospin independent.

To proceed, it is necessary to identify the dependence of the terms appearing in Eq.~(\ref{psum}) on the various single-particle and thermodynamic quantities.
Because $E^{\rm kin}_i$ depends only on $k$ and $\Sigma_S$, Eqs.~(\ref{fi}) and (\ref{fbari}) can also be written as
\begin{eqnarray}
 f_i &=& \frac1{1+\exp[(E^{\rm kin}_i(k,\Sigma_S) -\nu_i)/T]} \; ,  \\
 f_{\bar i} &=& \frac1{1+\exp[(E^{\rm kin}_i(k,\Sigma_S) + \nu_i)/T]} \; , 
\end{eqnarray}
with 
\begin{equation}
 \nu_i = \mu_i - \Sigma_V^i- \Sigma_R \label{nu} \; , 
\end{equation}
where $\nu_i$ is the so-called effective or kinetic chemical potential. Written in this way, one obtains Fermi-Dirac distribution functions equivalent to a noninteracting system with effective chemical potentials $\nu_i$ and particle masses $m_i^*$. The kinetic pressure of nucleon $i$ thus depends only on $T$, $\nu_i$, and $\Sigma_S$:
\begin{eqnarray}
 P^{\rm kin}&=&\sum_i P^{\rm kin}_i (T,\nu_i,\Sigma_S) \; . \label{pkin}
\end{eqnarray}
The interaction pressure is only a function of the self-energies, 
\begin{eqnarray}
P^{\rm int}=P^{\rm int}(n_B,\Sigma_S, \boldsymbol \Sigma_V) \; , \label{pint}
\end{eqnarray}
where $\boldsymbol \Sigma_V = \{\Sigma_V^i\}$ and has no direct dependence on temperature or the chemical potentials, as shown below. Furthermore, the dependence of $P^{\rm int}$ on the baryon number density $n_B$, defined as
\begin{eqnarray}
 n_B&=&\sum_i n_i \label{defnb} \\ 
    &=& n_n+n_p\; , 
\end{eqnarray}
is only present in density-dependent models. This follows from the following relations for the rearrangement contributions that are based on thermodynamic consistency:
\begin{eqnarray}
P^{\rm R}&=&n_B\Sigma_R  \label{pr} \\
\Sigma_R&=&-\left.\frac{\partial P^{\rm int}}{\partial n_B}\right|_{\Sigma_S, \boldsymbol \Sigma_V} \; . \label{sigmar}
\end{eqnarray}
By using Eq.~(\ref{sigmar}) in Eq.~(\ref{pr}) and because of Eq.~(\ref{pint}), one also has
\begin{eqnarray}
P^{\rm R}=P^{\rm R}(n_B,\Sigma_S, \boldsymbol \Sigma_V) \; . \label{prdep}
\end{eqnarray}
Note that $n_B$ that appears in the expressions above eventually is also a function of $T$ and $\mu_i$ and has to be determined in a self-consistent solution. 

In meson-exchange models for the nucleon interactions, the self-energies $\Sigma_S$ and $\boldsymbol \Sigma_V$ are actually fixed by the corresponding equations of motion of the meson fields, which is used in the following. The equations of motion are given in the implicit form
\begin{eqnarray}
 0&=&\left.\frac{\partial P}{\partial \Sigma_V^i}\right|_{T,\boldsymbol \mu, \Sigma_S, \Sigma_V^{j \neq i}} \; ,\label{ueom} \\
 0&=&\left.\frac{\partial P}{\partial \Sigma_S}\right|_{T,\boldsymbol \mu, \boldsymbol \Sigma_V} \; , \label{seom}
\end{eqnarray}
with $\boldsymbol \mu = \{\mu^i\}$. These equations extremize the grand-canonical potential. Because momentum-independent interactions are considered, the equilibrium values of the self-energies $\Sigma_S$ and $\Sigma_V^i$ are thus functions of only $T$ and the chemical potentials $\mu_n$ and $\mu_p$, $\Sigma_S= \Sigma_S(T,\mu_n,\mu_p)$, and $\Sigma_V^i= \Sigma_V^i(T,\mu_n,\mu_p)$. Note that only $\Sigma_S$ and $\Sigma_V^i$ appear in Eqs.~(\ref{ueom}) and (\ref{seom}), but not $\Sigma_R$, which illustrates the different role of the rearrangement part of the self-energies. 

The net number densities $n_i$, i.e., the difference between nucleon and antinucleon number densities, are defined in the usual way as
\begin{eqnarray}
 n_i=\left.\frac{d P}{d \mu_i}\right|_{T,\mu_{j\neq i}} \; . \label{eq_defni}
\end{eqnarray}
Here the notation was introduced to use ``$d$'' instead of ``$\partial$'' for partial derivatives, where only the other thermodynamic variables, but not the values of the self-energies, are kept constant. Thus, derivatives with $d$ are standard thermodynamic derivatives and include the changes of the fields, e.g., $\frac{d\Sigma_S}{d\mu_i} \frac{\partial P}{\partial \Sigma_S}$ . 

Using Eqs.~(\ref{pkin})--(\ref{seom}), from Eq.~(\ref{eq_defni}) one obtains
\begin{eqnarray}
 n_i&=&\left.\frac{\partial P^{\rm kin}}{\partial \nu_i}\right|_{T,\nu_{j\neq i},\Sigma_S} \label{eq_defni_noint} \\
&=&n_i^{\rm kin}(T,\nu_i,\Sigma_S)\label{ni2} \\
&=& \frac1{\pi^2} \int_0^\infty dk~k^2 (f_i-f_{\bar i}) \; . 
\end{eqnarray}
The interacting system still obeys Fermi-Dirac statistics, and obviously the interactions should not contribute to the particle numbers. Therefore, the densities $n_i$, defined by Eq.~(\ref{eq_defni}), have to be equal to the ones obtained only from the kinetic pressure for noninteracting particles with the same effective mass, respectively self-energy $\Sigma_S$, as expressed in the three preceeding equations. On the contrary, if the interaction part had a direct dependence on $\mu_i$ these relations would have been violated. This explains why $\mu_i$ does not appear in the functional dependence of $P^{\rm int}$; see Eq.~(\ref{pint}).

In RMF models, the fields behave like classical fields, and thus they do neither contribute to the entropy of the system. Instead, the entropy is just given by the kinetic contribution of nucleons,
\begin{eqnarray}
 s&=&-\left.\frac{d P}{d T}\right|_{\boldsymbol \mu}  \\
&=& s^{\rm kin} \; ,
\end{eqnarray}
with 
\begin{eqnarray}
s^{\rm kin} &=& -\left.\frac{\partial P^{\rm kin}}{\partial T}\right|_{\boldsymbol \nu, \Sigma_S}  \; .
\end{eqnarray}
If one uses the equations of motion (\ref{ueom}) and (\ref{seom}), this directly implies that $P^{\rm int}$ cannot have a direct temperature dependence and thus justifies Eq.~(\ref{pint}).

Using this information, for the internal energy density one finds
\begin{eqnarray}
 \epsilon&=&Ts -P +\sum_i n_i \mu_i \\
&=& \epsilon^{\rm kin} + \epsilon^{\rm int}  \; ,
\end{eqnarray}
whereas 
\begin{eqnarray}
\epsilon^{\rm kin} &=& Ts^{\rm kin} - P^{\rm kin} +\sum_i n_i \nu_i \; , \label{ekin} \\
\epsilon^{\rm int} &=& - P^{\rm int} -P^{\rm R}+ \sum_i n_i (\Sigma_V^i + \Sigma_R)\;  ,
\end{eqnarray}
and with Eqs.~(\ref{pr}) and (\ref{defnb}) this leads to
\begin{eqnarray}
    \epsilon^{\rm int} &=& - P^{\rm int} + \sum_i n_i \Sigma_V^i . \label{eint}
\end{eqnarray}
The rearrangement terms do not appear here; they cancel each other in the interaction part of the internal energy density. 

This leads to the free-energy density
\begin{eqnarray}
 f &=& \epsilon - Ts \\
&=& f^{\rm kin}+f^{\rm int} \; ,
\end{eqnarray}
whereas 
\begin{eqnarray}
f^{\rm kin} &=& \epsilon^{\rm kin} -Ts^{\rm kin}  \\
 &=& - P^{\rm kin} +\sum_i n_i \nu_i \; ,\label{fkin} \\
f^{\rm int} &\equiv&  \epsilon^{\rm int}\; . \label{efequiv}
\end{eqnarray}
Because there is no contribution of the interactions to the entropy, the interaction part of the free energy is identical to the interaction part of the internal energy. Thus, in the following, only $\epsilon^{\rm int}$ is used instead of $f^{\rm int}$.

To proceed, it is helpful to change to an equivalent canonical formulation, where the particle number densities $n_i$ and the temperature $T$ are used as state variables. The kinetic free-energy density has no direct dependence on the vector self-energies, as can be seen from Eqs.~(\ref{fkin}), (\ref{pkin}), and (\ref{ni2}). Because of Eqs.~(\ref{eint}) and (\ref{pint}), the interaction part has a direct dependence on the densities and the self-energies, but not on temperature. Thus, one can write
\begin{eqnarray}
 f=f^{\rm kin}(T,\boldsymbol n, \Sigma_S)+\epsilon^{\rm int}(\boldsymbol n, \Sigma_S, \boldsymbol \Sigma_V) \; ,
\end{eqnarray}
with $\boldsymbol n = \{n_i\}$. The equivalent equations of motion to Eqs.~(\ref{ueom}) and (\ref{seom}) in the canonical formulation are
\begin{eqnarray}
 0&=&\left.\frac{\partial f}{\partial \Sigma_V^i}\right|_{T,\boldsymbol n, \Sigma_S, \Sigma_V^{j \neq i}} \; ,\label{ueomf} \\
 0&=&\left.\frac{\partial f}{\partial \Sigma_S}\right|_{T,\boldsymbol n, \boldsymbol \Sigma_V} \; . \label{seomf} 
\end{eqnarray}
These equations of motion represent implicit equations which fix $\Sigma_S=\Sigma_S(T,\boldsymbol n)$ and $\Sigma_V^i=\Sigma_V^i(T,\boldsymbol n)$. The relations analogous to Eqs.~(\ref{eq_defni}) and (\ref{eq_defni_noint}) read
\begin{eqnarray}
 \mu_i &=& \left.\frac{d f}{d n_i}\right|_{T, n_{j \neq i}} \label{mui}
\end{eqnarray}
and 
\begin{eqnarray}
 \nu_i &=& \left.\frac{\partial f^{\rm kin}}{\partial n_i}\right|_{T, n_{j \neq i},\Sigma_S} \; . \label{nui}
\end{eqnarray}
 
Note that Eqs.~(\ref{eint}), (\ref{pint}), and (\ref{sigmar}) imply 
\begin{eqnarray}
 \left.\frac{\partial \epsilon^{\rm int}}{\partial n_i}\right|_{n_{j \neq i},\Sigma_S,\boldsymbol \Sigma_V} &=& \Sigma_R + \Sigma_V^i \; . \label{ui2}
\end{eqnarray}
Equations~(\ref{mui}) and (\ref{nui}) are consistent with Eq.~(\ref{nu}), which can be shown easily by making use of the last equation and the equations of motion (\ref{ueomf}) and (\ref{seomf}). 

Next, one introduces $\Delta \Sigma_V$, which replaces $\Delta U$ in the covariant formulation. It is defined as 
\begin{equation}
\Delta \Sigma_V = \Sigma_V^{n} - \Sigma_V^{p}  \; , \label{deltazv}
\end{equation}
with
\begin{equation}
\Delta \Sigma_{VR} = \Sigma_{VR}^{n} - \Sigma_{VR}^{p}  \; . \label{deltazvtot}
\end{equation}
Comparing with Eq.~(\ref{sigmavr}), one has
\begin{equation}
\Delta \Sigma_{VR} =\Delta \Sigma_V   \;  .
\end{equation}
The difference of the total RMF vector potentials is equal to the difference of the vector potentials without the rearrangement terms. Thus, in the following only $\Delta \Sigma_V$ is used.

Note that in the nonrelativistic case, when $k\ll m^*_i$, the single-particle energies can be approximated as
\begin{eqnarray}
 E_i&\simeq& m_i + \frac{k^2}{2 m^*_i}+\Sigma_S+\Sigma_V^i+\Sigma_R \; . \label{e_isim}
\end{eqnarray}
Accordingly, one can define the approximated, nonrelativistic mean-field potentials $U_i$,
\begin{eqnarray}
 U_i&=& \Sigma_S+\Sigma_V^i+\Sigma_R \; , \label{defui}
\end{eqnarray}
and their difference,
\begin{eqnarray}
\Delta U = U_n-U_p \; . \label{defdu}
\end{eqnarray}
Because the scalar self-energies of neutrons and protons are the same, here one has
\begin{eqnarray}
\Delta U = \Delta \Sigma_V \; . \label{dudz}
\end{eqnarray}
By using the definitions of the baryon number density $n_B=n_n+n_p$ and the proton fraction $Y_p=n_p/n_B$, and Eq.~(\ref{ui2}), $\Delta \Sigma_V$ can be written as
\begin{eqnarray}
 \Delta \Sigma_V &=& -\frac1{n_B}  \left.\frac{\partial \epsilon^{\rm int}}{\partial Y_p}\right|_{n_B,\Sigma_S, \boldsymbol \Sigma_V}  \; . \label{deltazv2}
\end{eqnarray}
This is a very intuitive expression: The change of the interaction part of the energy with changing asymmetry at fixed self-energies is given by the potential difference of neutrons and protons.

Here is a brief summary of what has been achieved so far: It is clear that the full knowledge of the vector and scalar self-energies, either as a function of temperature and the chemical potentials or of temperature and densities, provides the full information about the EOS, i.e., of all thermodynamic quantities. The functional dependence of these thermodynamic quantities on the state variables and the self-energies was derived. This is useful below for connecting $\Delta \Sigma_V$ with the potential symmetry energy. Note again that all equations presented in this section obey the standard rules of thermodynamic consistency, because they have been derived consistently from the grand-canonical potential.

\subsection{Approximating the asymmetry dependence}
Next, the approximation for the isospin dependence at finite temperature will be discussed for the generic mean-field model as specified above. The symmetry energy will be introduced and a derivation for its relation to $\Delta \Sigma_V$ will be given. 

For cold nucleonic matter, the EOS is well approximated by a parabolic expansion in terms of the asymmetry parameter $\delta$, 
\begin{equation}
\delta = 1-2Y_p \; , 
\end{equation}
around $\delta=0$. However, even if the interactions are completely isospin symmetric, the mass splitting of neutrons and protons $Q$ leads to a significant isospin-symmetry breaking of the EOS, especially relevant at low density. As a consequence, the proton fraction of the minimum of the thermodynamic potential (including the rest masses) is generally larger than $0.5$ and its value is temperature and density dependent. Nevertheless, one can expand the EOS around $Y_p^0=0.5$, respectively $\delta=0$, if one also includes a linear term. Therefore, one should consider the following expansion of the free energy per baryon $F=f/n_B$ with the rest-mass splitting included:
\begin{eqnarray}
 F&=&F(T,n_B,Y_p^0) \nonumber \\
&&+\delta F_{\rm lin}(T,n_B)+ \delta^2 F_{\rm sym}(T,n_B)+ {\cal O} (\delta^3) \; . \label{f}
\end{eqnarray}
The coefficient of the linear term of the expansion of $F$ is defined as
\begin{eqnarray}
 F_{\rm lin}&=& \left.\frac{d F}{d \delta}\right|_{T, n_B, \delta=0} \\
            &=& -\frac12 \left.\frac{d F}{d Y_p}\right|_{T, n_B, Y_p=Y_p^0} \\
            &=&-\frac12 F'(T,n_B,Y_p^0)  \; , 
\end{eqnarray}
and of the quadratic term as
\begin{eqnarray}
 F_{\rm sym}&=& \frac12 \left.\frac{d^2 F}{d \delta^2}\right|_{T, n_B, \delta=0} \\
            &=& \frac18 \left.\frac{d^2 F}{d Y_p^2}\right|_{T, n_B, Y_p=Y_p^0} \\
            &=& \frac18 F''(T,n_B, Y_p^0) \; ,
\end{eqnarray}
which is the free symmetry energy.

By using the equations of motion (\ref{ueomf}) and (\ref{seomf}), one obtains for the first derivative
\begin{eqnarray}
F'&=&\left.\frac{d F}{d Y_p}\right|_{T, n_B} = \left.\frac{\partial F}{\partial Y_p}\right|_{T, n_B, \Sigma_S, \boldsymbol \Sigma_V} \label{f'} \\
&=& \nu_p-\nu_n-\Delta \Sigma_V \; ,
\end{eqnarray}
and thus
\begin{eqnarray}
 F_{\rm lin}&=& \frac12 \left(\nu_n(T,n_B,Y_p^0)-\nu_p(T,n_B,Y_p^0)\right) \; , \label{flin}
\end{eqnarray}
where it was used that $\Delta \Sigma_V (T,n_B,Y_p=0.5)=0$ for all the models that are considered here. This is because exact isospin-symmetry is incorporated in the interactions. Note that for $m_n=m_p$ Eq.~(\ref{flin}) would equal to zero; i.e., the linear term would be absent.

For the second derivative one has
\begin{eqnarray}
F''&=&\left.\frac{d^2 F}{d Y_p^2}\right|_{T, n_B} \\
&=& n_B\left(\left.\frac{\partial \nu_p}{\partial n_p}\right|_{T, \Sigma_S}+\left.\frac{\partial \nu_n}{\partial n_n}\right|_{T, \Sigma_S}\right) \nonumber \\
&& + \left.\frac{d \Sigma_S}{d Y_p}\right|_{T, n_B} 
\left.\frac{\partial (\nu_n-\nu_p)}{\partial \Sigma_S}\right|_{T, n_B,Y_p}  \nonumber \\
&& -\left.\frac{d \Delta \Sigma_V}{d Y_p}\right|_{T, n_B} \label{f''}\; .
\end{eqnarray}

The first line of Eq.~(\ref{f''}) is the direct kinetic contribution to the free symmetry energy. The second line comes from the dependence of the scalar self-energy in the kinetic energy on asymmetry. Even though it depends on the scalar interactions, it is accounted as a kinetic term, because it originates from $F^{\rm kin}=f^{\rm kin}/n_B$. Note that for $m_n=m_p$, this second term would be zero for $Y_p=Y_p^0=0.5$. Thus, the kinetic free symmetry energy is defined to be
\begin{eqnarray}
F^{\rm kin}_{\rm sym}&=& \frac18 n_B\left(\left.\frac{\partial \nu_p}{\partial n_p}\right|_{T, \Sigma_S}+\left.\frac{\partial \nu_n}{\partial n_n}\right|_{T, \Sigma_S}\right) \nonumber \\
&& +\frac18 \left.\frac{d \Sigma_S}{d Y_p}\right|_{T, n_B} \left.\frac{\partial (\nu_n-\nu_p)}{\partial \Sigma_S}\right|_{T, n_B,Y_p}  \label{fkinsym} \; , 
\end{eqnarray}
and, correspondingly, the interaction symmetry energy is
\begin{eqnarray}
E^{\rm int}_{\rm sym}&=& -\frac18\left.\frac{d \Delta \Sigma_V}{d Y_p}\right|_{T, n_B} \; , \label{fintsymm}
\end{eqnarray}
both evaluated at $Y_p^0$ and so that 
\begin{equation}
F_{\rm sym}(T,n_B)=F^{\rm kin}_{\rm sym}(T,n_B)+E^{\rm int}_{\rm sym}(T,n_B) \; . 
\end{equation}
Note that one can use the interaction symmetry energy instead of the interaction \textit{free} symmetry energy because this term originates from $f^{\rm int}$, which is identical to $\epsilon^{\rm int}$; see Eq.~(\ref{efequiv}). 

\begin{figure*}
\includegraphics[width=1.6\columnwidth]{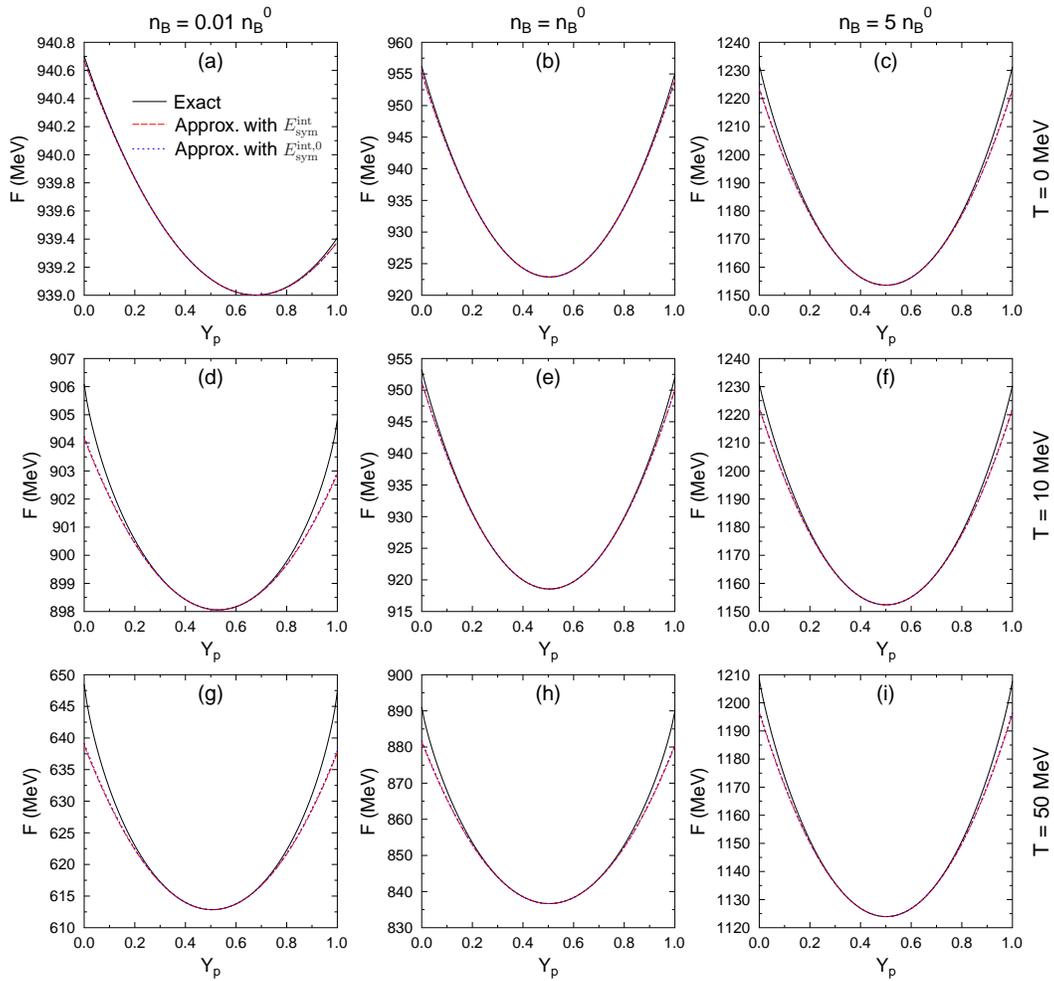}
\caption{(Color online) Free energy per baryon (black solid lines) and two approximations of it based on $E^{\rm int}_{\rm sym}$ (red dashed lines) and $E^{\rm int,0}_{\rm sym}$ (blue dotted lines), as a function of the proton fraction, calculated with the density-dependent RMF model DD2 \cite{typel09}. The columns show results for densities of 0.01~$n_B^0$, $n_B^0$, and $5n_B^0$ (from left to right), the rows for temperatures of 0, 10, and 50 MeV (from top to bottom).}
\label{fig:dd2_f}
\end{figure*}

\begin{figure*}
\includegraphics[width=1.6\columnwidth]{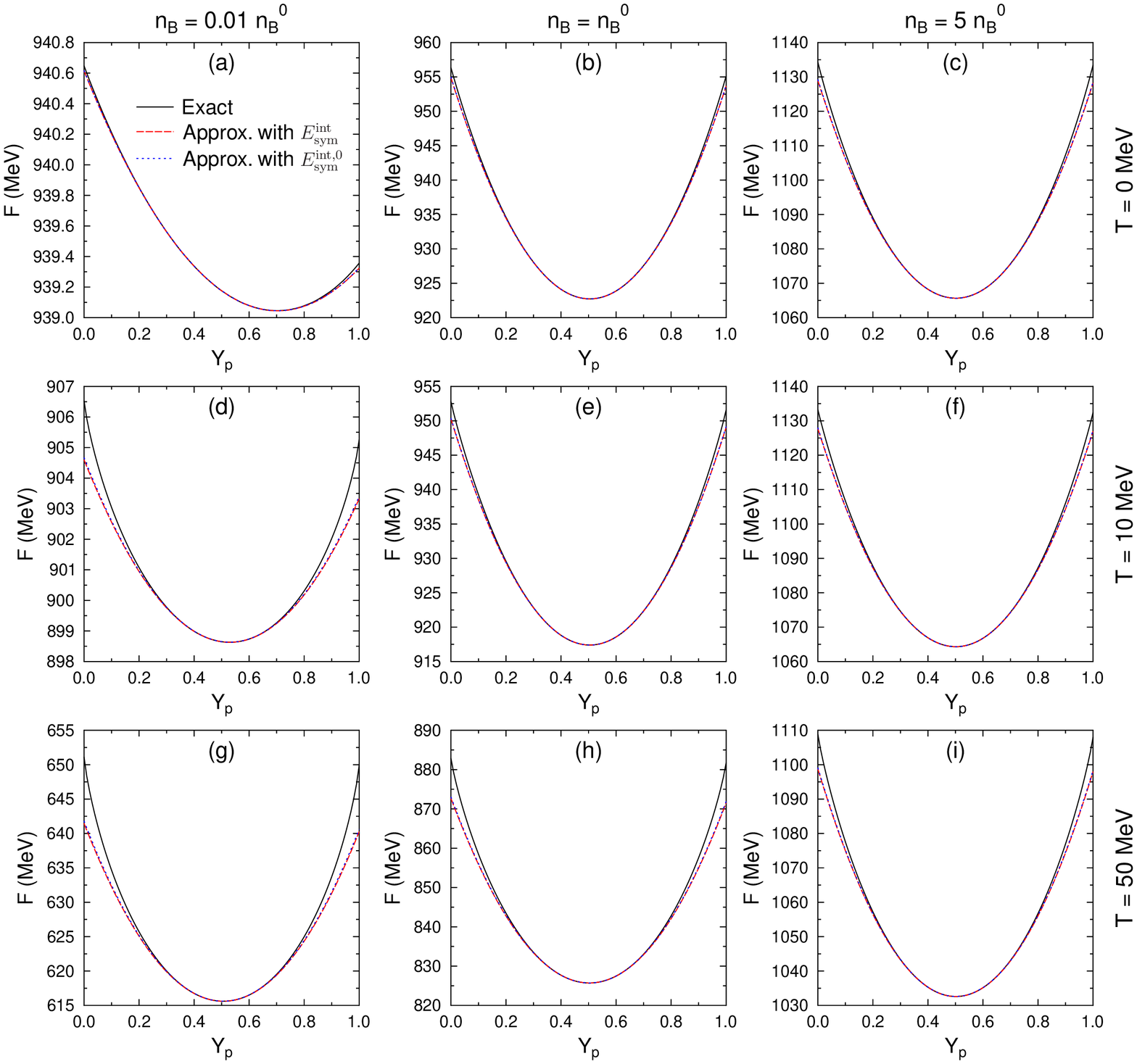}
\caption{(Color online) As Fig.~\ref{fig:dd2_f}, but for the nonlinear RMF model SFHo \cite{steiner13}.}
\label{fig:sfho_f}
\end{figure*}

Next one also expands $\Delta \Sigma_V$ in $Y_p$ around $Y_p^0$:
\begin{eqnarray}
 \Delta \Sigma_V= (Y_p-Y_p^0) \left.\frac{d \Delta \Sigma_V}{d Y_p}\right|_{T, n_B,Y_p^0}+{\cal O} (\Delta Y_p^2) \; .
\end{eqnarray}
This leads to: 
\begin{eqnarray}
 \Delta \Sigma_V = 4 (1-2Y_p) E^{\rm int}_{\rm sym}+{\cal O} (\Delta Y_p^2) \; . \label{yeah}
\end{eqnarray}
This expression is an important result of the present investigation. It shows that $\Delta \Sigma_V$, and thus also the nonrelativistic potential difference $\Delta U$, is given by the potential or interaction part of the symmetry energy, up to linear order in $Y_p$. Note that, in this order, the definition of $E^{\rm int}_{\rm sym}$ is equivalent to the Lane potential \cite{lane62} modulus factor $8$.

$E^{\rm int}_{\rm sym}$ can also be expressed as
\begin{eqnarray}
E^{\rm int}_{\rm sym}&=& \frac18\left.\frac{d }{d Y_p}\right|_{T, n_B} \left.\frac{\partial E^{\rm int} }{\partial Y_p}\right|_{n_B, \Sigma_S, \boldsymbol \Sigma_V}\; ,
\end{eqnarray}
with $E^{\rm int}= \epsilon^{\rm int}/n_B$, as is obvious by comparing with Eqs.~(\ref{deltazv2}) and (\ref{fintsymm}).
It should be emphasized that 
\begin{eqnarray}
E^{\rm int}_{\rm sym}\neq \frac18 \left.\frac{d^2 E^{\rm int}}{d Y_p^2}\right|_{T, n_B} \; .
\end{eqnarray}
This means that if one would make an expansion of $E^{\rm int}$ in terms of asymmetry, additional terms to the one with $E^{\rm int}_{\rm sym}$ would be present. These do not show up in the expansion (\ref{f}) of $F$, because they cancel with terms coming $F^{\rm kin}$. The cancellation is caused by the equations of motion.

The proposed decomposition of $F_{\rm sym}$ into a kinetic and an interaction part, where the latter is given by the vector self-energy contribution, is equivalent to what was reported in Ref.~\cite{cai12}. In this article, the nuclear symmetry energy and its slope parameter $L$ were decomposed in terms of the Lorentz covariant nucleon self-energies, using the Hugenholtz-Van Hove theorem at zero temperature. In Ref.~\cite{cai12}, also momentum-dependent interactions and a scalar isovector interaction were considered, which are not taken into account here. However, the derivation of Ref.~\cite{cai12} is only valid for $T=0$, whereas the present results are for arbitrary temperature.

In most RMF models, the vector self-energies do not depend on temperature. In the eight models that are considered, only for SFHo and SFHx they do have a temperature dependence owing to a coupling of the scalar meson with the vector mesons. However, even for these two models the temperature dependence of $\Sigma_V^i$ is only very weak. Consequently, Eq.~(\ref{yeah}) suggests that one could also use $E^{\rm int,0}_{\rm sym}(n_B):=E^{\rm int}_{\rm sym}(T=0,n_B)$ in the expansion of $F$ and in the relation to $\Delta \Sigma_V$ instead of $E^{\rm int}_{\rm sym}(T,n_B)$. The performance of this further simplification where the interaction symmetry energy at zero temperature is used is examined below.

\subsection{Results}
\label{sec_results}
In Fig.~\ref{fig:dd2_f}, the free energy per baryon $F=f/n_B$ of the density-dependent RMF model DD2 \cite{typel09} is shown for various densities and temperatures by the black solid lines. In panel (a) it is clearly visible that the minimum of $F$ is obtained for $Y_p\sim0.7>0.5$, and that the EOS is not isospin symmetric around 0.5, because of the difference of the neutron and proton rest masses. For even lower densities, where $Q=m_n-m_p$ is the most important energy scale, these effects would be even more pronounced. 

The red dashed lines show the expansion of $F$ according to Eq.~(\ref{f}). For the blue dotted lines, $E^{\rm int}_{\rm sym}(T,n_B)$ has been replaced with $E^{\rm int,0}_{\rm sym}(n_B)$ in the expansion. Figure~\ref{fig:sfho_f} shows the same quantities, but for the nonlinear RMF model SFHo \cite{steiner13}. For DD2 it is confirmed that the two approximations give almost identical results. Also for SFHo, where the interaction symmetry energy has some temperature dependence, no notable differences occur. The temperature dependence of $E^{\rm int}_{\rm sym}$ is indeed negligible. In the comparison of the approximations with the exact results, one sees that higher-order terms become important for high asymmetries at high densities and/or high temperatures. Generally, the approximations underestimate $F$. 

\begin{figure*}
\includegraphics[width=1.6\columnwidth]{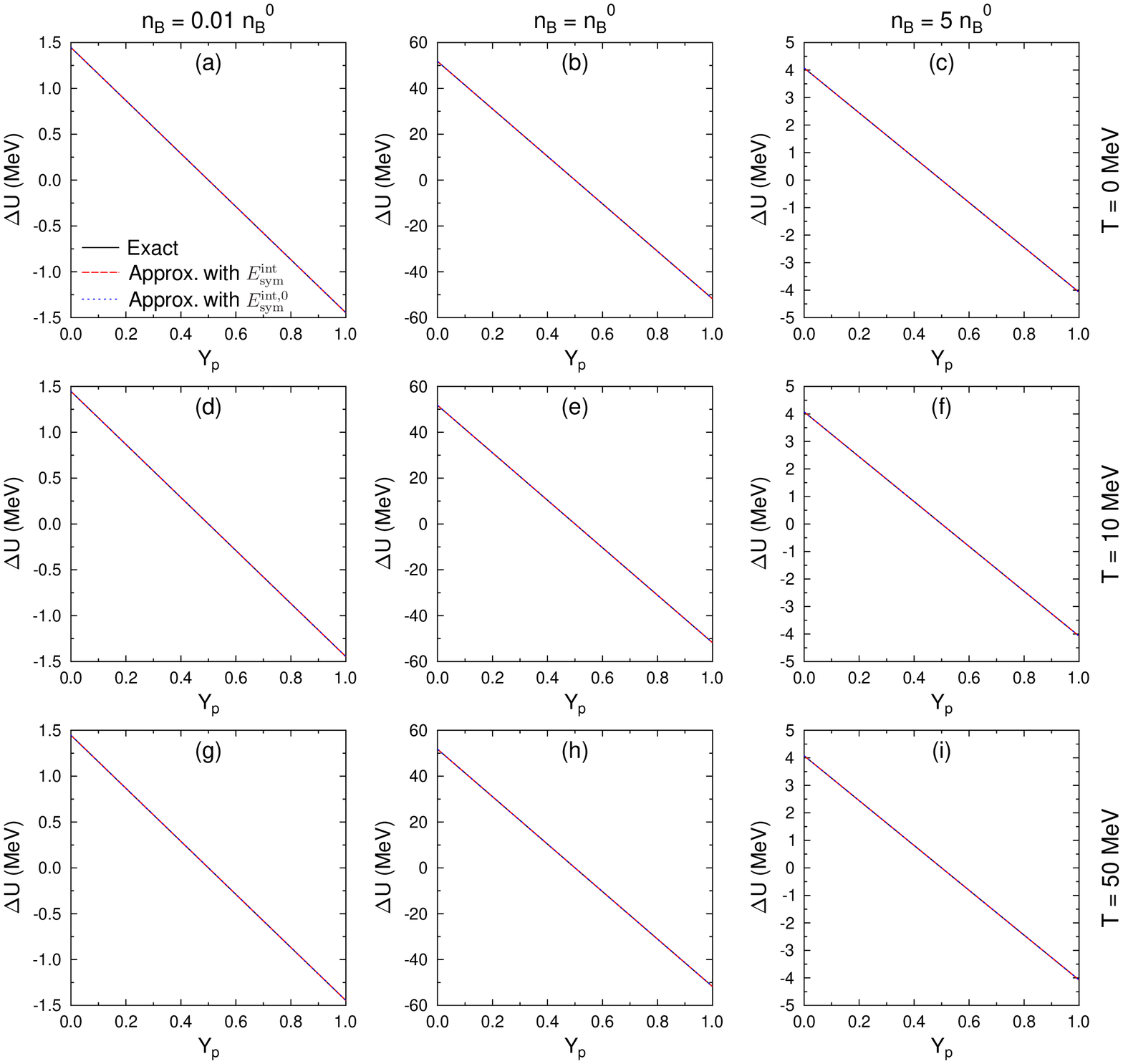}
\caption{(Color online) Potential difference $\Delta U$ (black lines) and two approximations of it based on $E^{\rm int}_{\rm sym}$ (red dashed lines) and $E^{\rm int,0}_{\rm sym}$ (blue dotted lines) as a function of the proton fraction, calculated with the density-dependent RMF model DD2 \cite{typel09}. The columns show results for densities of 0.01~$n_B^0$, $n_B^0$, and $5n_B^0$ (from left to right); the rows show results for temperatures of 0, 10, and 50 MeV (from top to bottom).}
\label{fig:dd2_du}
\end{figure*}
\begin{figure*}
\includegraphics[width=1.6\columnwidth]{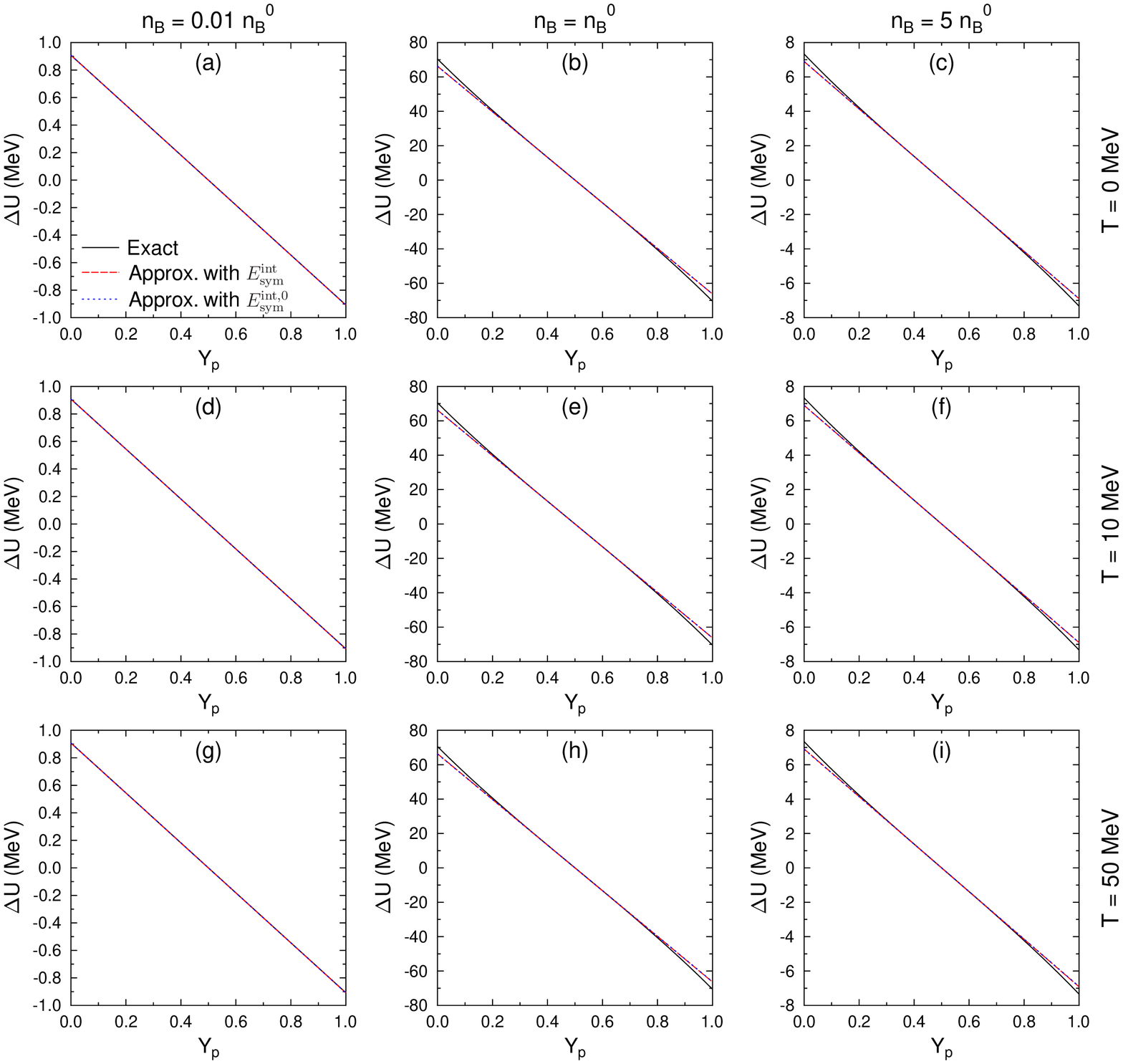}
\caption{(Color online) As Fig.~\ref{fig:dd2_du}, but for the nonlinear RMF model SFHo \cite{steiner13}.}
\label{fig:sfho_du}
\end{figure*}

The results for the difference of the vector self-energies $\Delta \Sigma_V$, respectively the potential difference $\Delta U$, are shown in Figs.~\ref{fig:dd2_du} and \ref{fig:sfho_du}, together with the two approximations based on $E^{\rm int}_{\rm sym}$ and $E^{\rm int,0}_{\rm sym}$ in Eq.~(\ref{yeah}). In DD2, the vector self-energies have a strictly linear dependence on asymmetry, because no cross-couplings between the different mesons are included. Furthermore, they are temperature independent, and indeed one can confirm that there are no notable differences to the exact calculation for both of the two approximations. This shows that the deviations found in Fig.~\ref{fig:dd2_f} between the exact calculation and the expansions originate from the kinetic free energy and the isospin dependence of the scalar self-energy. In Ref.~\cite{wellenhofer15}, a detailed analysis of the noninteracting contribution was given, and it was also found that the accuracy of the quadratic expansion decreases with increasing temperatures.

For $\Delta U$ of SFHo, shown in Fig.~\ref{fig:sfho_du}, one has both a temperature dependence and a nonlinear dependence on asymmetry of the vector self-energies owing to coupling of the vector isovector meson with other mesons. The deviations of the two approximations because of missing nonlinear terms are visible in Fig.~\ref{fig:sfho_du} for $n_B=n_B^0$ and $n_B=5n_B^0$. It should be noted that, overall, they are still small, and that the linear approximation works reasonably well. The temperature dependence, however, is so small that no differences are visible between the two approximations based on $E^{\rm int}_{\rm sym}(T,n_B)$ and $E^{\rm int,0}_{\rm sym}(n_B)$. 

\begin{figure*}
\includegraphics[width=1.6\columnwidth]{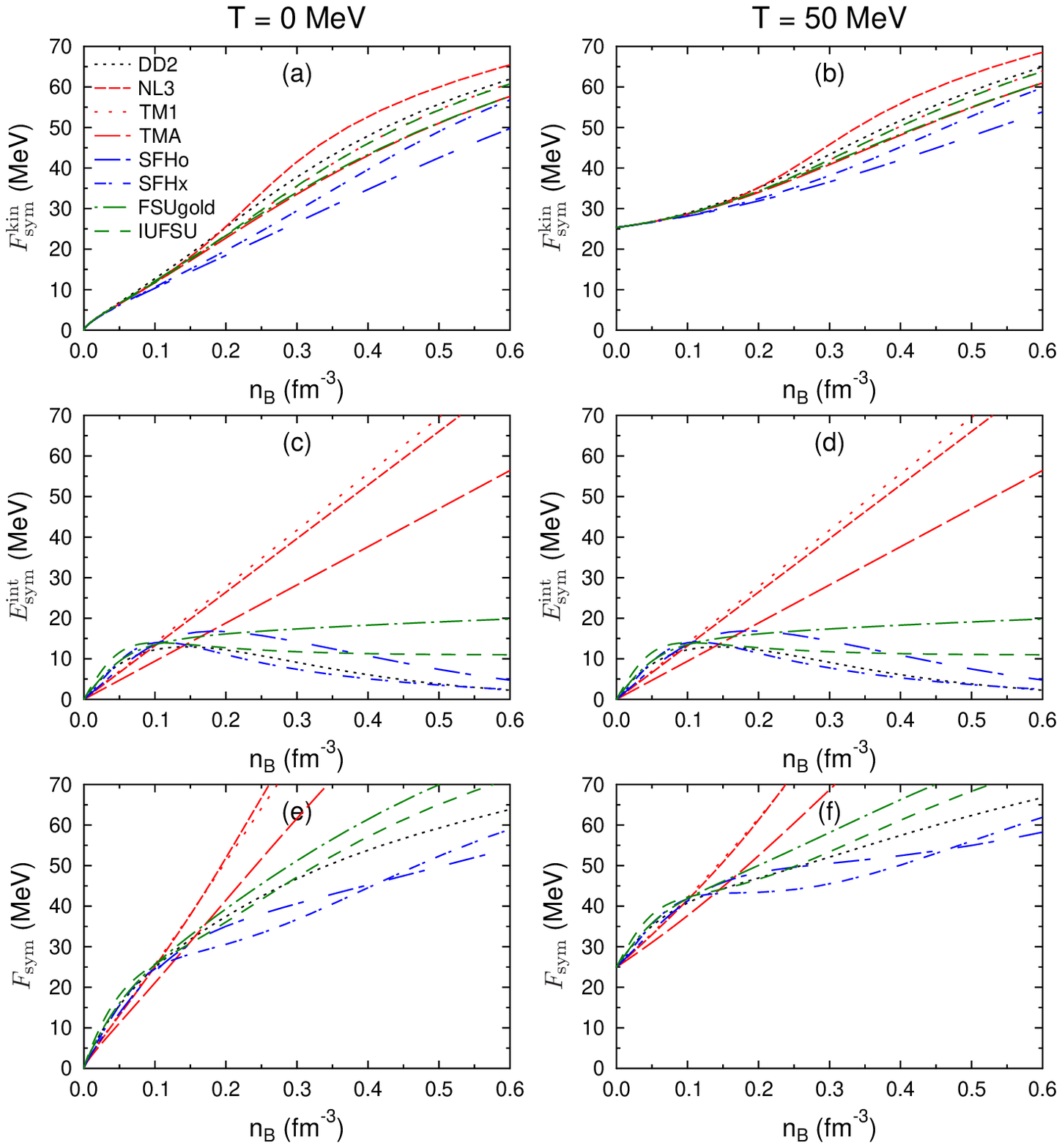}
\caption{(Color online) Top panels: Kinetic part of the free symmetry energy as a function of baryon number density for various RMF models. Middle panels: Interaction part of the symmetry energy. Bottom panels: Free symmetry energy. Left column is for $T=0$; the right column is for $T=50$~MeV.}
\label{fig:fsyms}
\end{figure*}
After having examined the asymmetry dependence, the density and temperature dependence of the free symmetry energy is discussed. Figure~\ref{fig:fsyms} shows the potential ($E_{\rm sym}^{\rm int}$) and kinetic part ($F_{\rm sym}^{\rm kin}$) of the free symmetry energy and its total value ($F_{\rm sym}$) calculated with the eight different RMF models for temperatures of 0 and 50~MeV. The density range shown extends to rather high densities, to cover also densities reached in cold NS, and to illustrate the overall behavior. As one can expect, $F_{\rm sym}^{\rm kin}$ has a strong temperature dependence. For $T=50$~MeV, even at zero density it keeps a high value, owing to the dependence of the entropy on asymmetry. Conversely, the temperature dependence of $E_{\rm sym}^{\rm int}$ is so small that is not visible in the figure by comparing panels (c) and (d).

At very low densities and high temperatures, the free symmetry energy is dominated by the kinetic contribution. However, if one compares the different RMF models, it is seen that the kinetic free symmetry energy is relatively similar for all of them, at least up to densities of $\sim 0.1$~fm$^{-3}$. From Eq.~(\ref{fkinsym}) it is obvious that $F_{\rm sym}^{\rm kin}$ is related to the scalar self-energy, and its dependence on density and asymmetry. The differences in the interaction part of the symmetry energy are significantly larger, and are visible in the total free symmetry energy already at 0.05~fm$^{-3}$. Above $n_B^0$, the different models give completely different results for both $E_{\rm sym}^{\rm int}$ and $F_{\rm sym}$, which illustrates the current ignorance of the symmetry energy at densities which are not reached in ordinary nuclei.

\begin{figure}
\includegraphics[width=0.85\columnwidth]{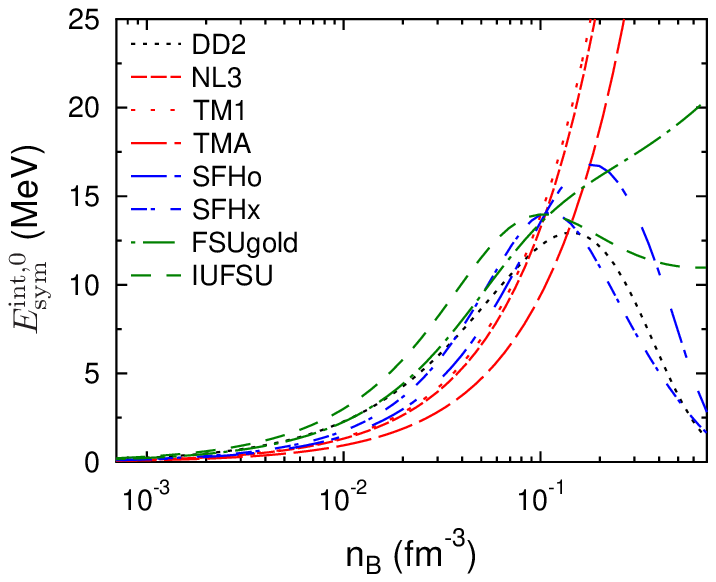}
\caption{(Color online) Interaction part of the symmetry energy at $T=0$ as a function of baryon number density for various RMF models.}
\label{fig:eintsym}
\end{figure}
In the following, for simplicity $E^{\rm int,0}_{\rm sym}(n_B)$ is used because it was shown above, that the temperature dependence of the interaction part of the symmetry energy is negligible. Figure~\ref{fig:eintsym} presents again the potential part of the symmetry energy at $T=0$ calculated with the eight different RMF models but restricted to the density range that is most relevant for envelopes of PNSs. Even below 0.1~fm$^{-3}$ there can be differences of more than 5~MeV. However, below it is shown that the low-density EOS is actually well constrained by nuclear experiments. It has a lower uncertainty than what is reflected here for the selection of theoretical models.

It is interesting to note, that the various models give very similar $E^{\rm int,0}_{\rm sym}$ around $0.1$~fm$^{-3}$. This is the density which is most relevant for properties of finite nuclei, which have been used in the fitting of the parameter sets in all of the models. At higher densities, the models diverge from each other. For example, in DD2, SFHo, and SFHx, the potential symmetry energy is approaching zero, whereas in the simple nonlinear models TM1, TMA, and NL3, it is increasing to extremely high values. 

The potential difference $\Delta \Sigma_V$, respectively $\Delta U$, is set not only by $E^{\rm int,0}_{\rm sym}$, but also by the asymmetry; see Eq.~(\ref{yeah}). The electron fraction in $\beta$ equilibrium, however, is determined by the free symmetry energy, i.e.\ the sum of the kinetic and interaction contribution. A high value of the free symmetry energy will lead to a lower asymmetry. In principle, this could lead to a compensation effect in Eq.~(\ref{yeah}) so that high symmetry energies would lead to lower values of  $\Delta U$. The electron fraction in $\beta$ equilibrium with charge neutrality ($Y_p=Y_e$) but without neutrinos is determined from the standard relation:
\begin{eqnarray}
 \mu_p+\mu_e = \mu_n \; .
\end{eqnarray}
In this equation one can use the expansion of Eq.~(\ref{f}) in the definition of the chemical potentials (\ref{mui}) to get
\begin{eqnarray}
 1-2Y_e \simeq (\mu_e-2F_{\rm lin})/4 F_{\rm sym} \; .
\end{eqnarray}
This is still an implicit equation to determine $Y_e$, because $\mu_e$ itself is a function of density, temperature, and asymmetry. Nevertheless, it shows that high values of $F_{\rm sym}$ drive the system to a more symmetric configuration.
  
To quantify the strength of the possible compensation effect, one can use the electron fraction in cold NSs, i.e., in $\beta$ equilibrium at zero temperature without neutrinos. This value of $Y_e$ corresponds to the final state of equilibrium where the newly born PNS will evolve to. At the onset of the collapse of the progenitor star, all EOSs will start with the same $Y_e$ profile. Differences in $Y_e$ in the subsequent evolution will emerge owing to different rates and/or different final equilibria. Therefore, the largest differences in $Y_e$ for different EOSs can be expected for cold NS. Furthermore, the $\beta$-equilibrium $Y_e$ can be seen as a general lower bound for $Y_e$, which in turn gives the highest values of $\Delta U$.

\begin{figure}
\includegraphics[width=0.85\columnwidth]{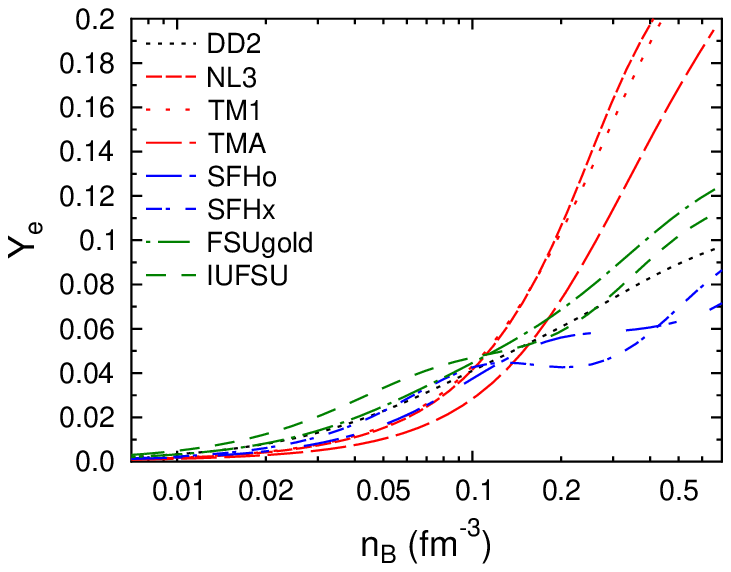}
\caption{(Color online) Electron fraction in $\beta$ equilibrium and $T=0$ as a function of baryon number density for various RMF models.}
\label{fig:yp}
\end{figure}
\begin{figure}
\includegraphics[width=0.85\columnwidth]{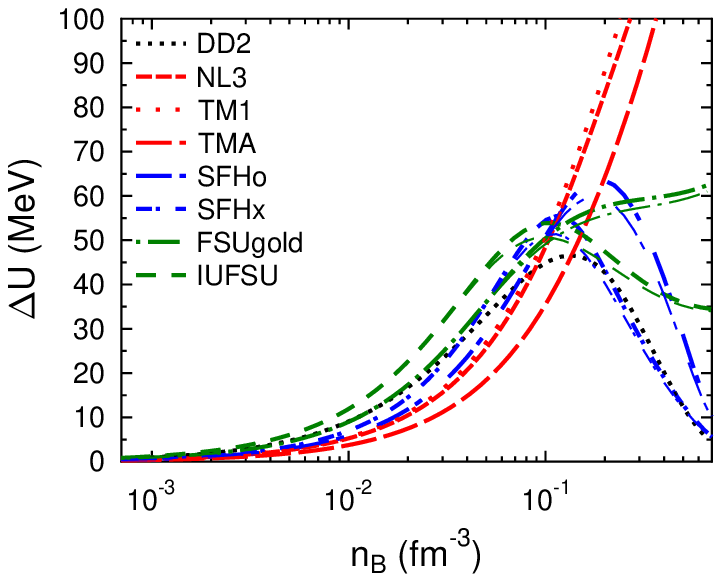}
\caption{(Color online) Vector potential difference for cold NSs as a function of baryon number density for various RMF models. Thick lines, exact calculation; thin lines, approximation by $E^{\rm int,0}_{\rm sym}$.}
\label{fig:du}
\end{figure}
Figure~\ref{fig:yp} shows the electron fraction for conditions of cold NSs for the various models. Some substantial variation is found, especially at high densities, reflecting the different symmetry energies. The thick lines in Fig.~\ref{fig:du} show the corresponding values of $\Delta U$, respectively $\Delta \Sigma_V$. By comparing with Fig.~\ref{fig:eintsym}, one sees that the compensation effect of the different electron fractions is not very important at low densities. Also at high densities, the qualitative behavior of $\Delta U$ is still very similar to $E^{\rm int,0}_{\rm sym}$. If one considers that matter has only a low asymmetry at the progenitor stage, leading to vanishingly small values of $\Delta U$, one can conclude that $\Delta U$ will evolve in the supernova from $\sim 0$ to the values shown in Fig.~\ref{fig:du}.  

The thin lines in Fig.~\ref{fig:du} show the results for the approximation 
\begin{equation}
\Delta U \simeq 4 (1-2Y_p) E^{\rm int,0}_{\rm sym} \; , \label{du_approx}
\end{equation}
i.e., using only the linear expansion and the interaction part of the symmetry energy at zero temperature. 
For TM1, TMA, NL3, and DD2 no deviations are visible. Only for the models SFHo, SFHx, IUFSU, and FSUgold deviations are found compared to the exact results, which can be attributed to nonlinear terms in $Y_e$. They arise in these models because the vector isovector meson is coupled with other mesons, as mentioned before. The deviations from strictly linear behavior are also visible in Fig.~\ref{fig:sfho_du}, e.g., in panel (b). Nevertheless, the approximation of Eq.~(\ref{du_approx}) still reproduces the overall behavior quite well, especially if one takes into account the extremely low values of $Y_e$ occurring here (compare with Fig.~\ref{fig:yp}).

\subsection{Experimental Constraints}
\label{sec_exp}
\begin{figure}
\includegraphics[width=0.85\columnwidth]{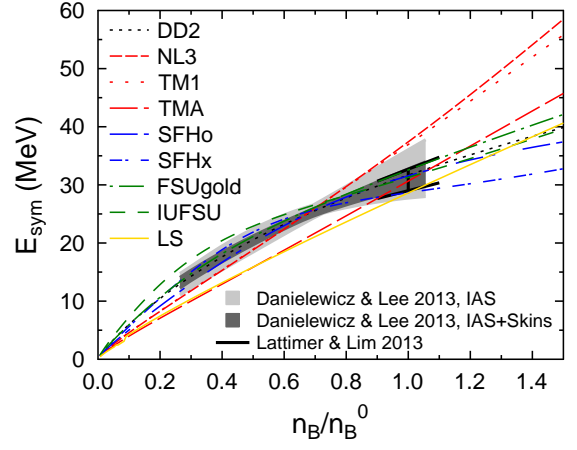}
\caption{(Color online) Symmetry energy at $T=0$ for the RMF models in comparison with the constraints from Lattimer and Lim \cite{lattimer2013} and from Danielewicz and Lee \cite{danielewicz2013}. ``LS'' shows the symmetry energy for the EOS of Lattimer and Swesty \cite{lattimer91}.}
\label{fig:esym}
\end{figure}
Figure~\ref{fig:esym} shows the symmetry energy at $T=0$ together with experimental constraints. The results shown for the RMF models are identical to the data presented in Fig.~\ref{fig:fsyms} (e). In addition, the symmetry energy of the SN EOS of Lattimer \& Swesty \cite{lattimer91} is included. This SN EOS model is frequently used in CCSN simulations, and EOS routines are available for three different values of the nuclear incompressibility. The symmetry energy is the same for all of the three variants. 

The gray shaded regions are taken from Ref.~\cite{danielewicz2013} where isobaric analog states (IAS) were used to extract the density-dependent symmetry energy. The covered density range corresponds to the conditions probed in finite nuclei. Note that the smallest uncertainty is obtained around 0.75~$n_B^0$ with $E_{\rm  sym}\sim 25$~MeV, which can be considered as an average density in nuclei. The dark gray shaded region (taken from the same reference) utilizes results for neutron skin thicknesses in addition, which puts constraints on the slope parameter $L$ of the symmetry energy at nuclear saturation density. This tightens the constraints significantly if combined with the analysis of IAS. The dark lines are the final results from Lattimer and Lim \cite{lattimer2013}, who provide a compilation of various different theoretical, experimental, and observational constraints for the value of the symmetry energy at $n_B^0$, $J$, and the slope parameter $L$. The vertical line represents the allowed region in $J$ and the two 
diagonal lines represent the allowed slope. It is seen that the constraints of Refs.~\cite{danielewicz2013} and \cite{lattimer2013} agree very nicely. Note that a constraint region similar to the one from Ref.~\cite{danielewicz2013} was deduced earlier in Ref.~\cite{tsang2012} from heavy-ion collision experiments. The final results for $E_{\rm  sym}$ between 0.3 and 1~$n_B^0$ are consistent with the ones of Ref.~\cite{danielewicz2013}, but less stringent and therefore not shown here.

The qualitative differences observed in Fig.~\ref{fig:esym} for the different models can be related to the interaction terms which are included. The RMF models TM1, TMA, and NL3, which are based on a simple nonlinear Lagrangian with self-couplings of the scalar isoscalar meson and the vector isoscalar meson (only in TM1 and TMA), give a roughly linear density dependence of the symmetry energy, which is commonly known (cf., Ref.~\cite{dutra2014}). NL3 and TM1, which were directly fitted to nuclear binding energies, go through the $E_{\rm  sym}$ value of $\sim 25$~MeV at 0.75~$n_B^0$, but consequently their slope and value of $E_{\rm  sym}$ at $n_B^0$ is too high. TMA is based on an interpolation of two different parameter sets, and it is far away from the experimental constraints for $E_{\rm sym}$ below $n_B^0$. The symmetry energy of LS behaves also very linearly and is too low for $n_B\leq n_B^0$, and $L$ is too high. IUFSU and FSUgold are RMF models which include the cross coupling between the isoscalar and the isovector vector mesons. This introduces the necessary nonlinear 
dependence of the symmetry energy seen in the experimental data. Note, however, that $E_{\rm sym}$ of IUFSU below 0.7~$n_B^0$ is too high, 
even though it gives a good 
behavior around $n_B^0$. 

The two models SFHo and SFHx have been fitted to measurements of low NS radii \cite{steiner10,steiner13}. It is interesting to see that these two models give a good agreement with the experimental constraints. 
One can conclude that the subsaturation symmetry energy extracted from radius measurements of NSs is consistent with the experimental constraints of Refs.~\cite{danielewicz2013} and \cite{lattimer2013}.
SFHx, where the ``x'' abbreviates ``extreme,'' is called in this way because it gives an extremely soft symmetry energy, visible by the low slope, which is not reached in most other mean-field models. The required flexibility of the functionals of SFHo and SFHx is obtained by including various meson self- and cross interactions. DD2 is based on density-dependent coupling constants. Its prediction of the symmetry energy is right in the middle of the constraints shown in Fig.~\ref{fig:esym}. 

Note also that DD2 is the only model considered here, where the agreement does not imply nonlinear terms in $\Delta U$ (compare Figs.~\ref{fig:dd2_du} and \ref{fig:sfho_du}), corresponding to strong quartic terms in $F_{\rm int}$. Unfortunately, such higher-order terms are currently not well constrained. For a discussion of the fourth-order symmetry energy, see, e.g., Refs.~\cite{steiner06,cai2012}. Recently, there has been new work on this subject using perturbative chiral effective field theory (EFT) \cite{drischler14} and auxiliary field diffusion Monte Carlo \cite{gandolfi14}. Both of these works have shown that the quadratic expansion of cold nuclear matter works very well; however, they did not constrain the fourth-order symmetry energy directly. It would be very interesting to use either such \textit{ab initio} approaches or experiments to pin down the strength of a possible fourth-order symmetry energy coefficient which could be a useful guideline for developing new empirical density functionals.

In Ref.~\cite{fischer14}, the neutron-matter EOS of the same models as considered here were compared with results from chiral EFT (see also Ref.~\cite{krueger13}) and basically the same conclusions were drawn as above. The simple nonlinear models TM1, NL3, and in particular TMA, and also LS, provide too much binding for the neutron matter EOS at subsaturation densities. Furthermore, the neutron matter EOS of IUFSU has too high energies compared with chiral EFT, leading to high values of $E_{\rm sym}$. 

It was discussed above that the kinetic free symmetry energy shows only a small variation for the different RMF parametrizations below $n_B^0$. Related to this, in Ref.~\cite{chen2014} it was shown that the effective mass, which determines the kinetic free symmetry energy, is rather well constrained at saturation density by properties of finite nuclei in typical RMF models. This leads to important conclusions about the possible range for $\Delta U$. It permits to interpret the constraints for $E_{\rm sym}$ as constraints for $E_{\rm sym}^{\rm int}$ and thereby for $\Delta U$. The models LS, TMA, NL3, TM1, and IUFSU are not reliable at low densities, because their symmetry energies are outside of the gray band and their neutron matter EOSs are in strong disagreement with chiral EFT. Only the predictions of the models DD2, FSUgold, SFHo, and SFHx remain as reasonable candidates. Using these constraints, one thus obtains a more narrow band for $\Delta U$, spanned by FSUgold, DD2, SFHx, and SFHo. 

\subsection{Comparison with other works}
One could question if the temperature independence or, at least, very weak temperature dependence of the mean-field interactions that was found here is realistic. This is confirmed, e.g., by Ref.~\cite{fedoseew2014}, which shows that the temperature modifications of the nucleon vertices and nucleon self-energies is almost negligible, based on Dirac-Brueckner calculations. In Ref.~\cite{wellenhofer15}, chiral EFT was used to investigate asymmetric nuclear matter at finite temperatures. It was found that the interaction parts of the free symmetry energy and internal symmetry energy have only a moderate or even weak temperature dependence, which seems to be qualitatively similar to the present results.

It was shown in several theoretical works that correlations have an impact on the decomposition of the symmetry energy into the kinetic and potential contribution. It mostly originates from the tensor component of the nuclear force which induces the population of high-momentum states, see, e.g., Refs.~\cite{vidana2011,carbone2012,hen2014}. A significant reduction of the kinetic and a corresponding increase of the potential part is found. These effects, which are not present in the mean-field picture, are very interesting. However, the basic neutrino interaction rates that are presented below are not appropriate for such models. They should include effects of correlations in a consistent manner; see, e.g., Ref.~\cite{reddy99}. Regarding the effects of realistic nucleon-nucleon interactions on the neutrino emission in the wind phase of SN, therefore, a more detailed investigation would be required.

The nucleon potential difference and the ``nucleon symmetry potential'' were also calculated directly in many-body approaches employing realistic nucleon interactions, such as Brueckner-Hartree-Fock or Dirac-Brueckner; see, e.g., Refs.~\cite{zuo2005,zuo2006,vandalen2004,fuchs2006}. The impact of the symmetry potential on preequilibrium
nucleon emission in heavy-ion collisions was studied in Ref.~\cite{liu2014}. Also, experimental data for the nucleon optical potentials from nucleon-nucleus scattering experiments are available \cite{li2004}. In Ref.~\cite{antic2015}, such data were used to construct a new type of RMF interaction. The authors of Ref.~\cite{xu2014} used the optical model
analyses of proton-nucleus scattering data in a nonrelativistic framework, to investigate the implications on thermal properties of nuclear matter. In general, the different momentum dependence of the single-particle potentials, different effective mass splittings, and the usage of relativistic and nonrelativistic frameworks complicate the comparison with the results presented here. Further comparisons would be beyond the scope of the present investigation and are thus left for future study. 

In the recent work of Ref.~\cite{rrapaj2014}, the nucleon potential difference at finite temperature was calculated in the Hartree-Fock approximation for two different realistic interactions that fit measured scattering phase shifts. For a so-called ``pseudo potential'' a much larger potential difference was found compared to a chiral potential. This was explained by strong nonperturbative effects. Typical RMF models were found to lie in the band spanned by these two models, i.e., also giving lower values than the pseudo-potential. It will be interesting to see higher-order many-body calculations in the future which reduce the theoretical uncertainty and further constrain the mean-field models. In the same work, also the role of the deuteron bound-state contribution was evaluated. The possible error induced by not including the deuteron consistently were found to be smaller than the differences obtained from the two potentials. In Ref.~\cite{horowitz12}, the nucleon potential difference was investigated within the virial EOS. There the deuteron bound state was also included, but its role was not discussed any further. Also with this approach higher nucleon potential differences were observed than in typical RMF models. In Sec.~\ref{sec_snmatter}, a closer comparison with this work is given. 

\subsection{Elastic charged-current rates}
\label{sec_rates_rmf}
In this section simple expressions for the charged-current rates are specified which are based on the elastic \cite{bruenn85} and nonrelativistic approximations, but which take the mean-field effects into account. The final results are equivalent to what was reported in Refs.~\cite{reddy98,martinez12}. However, here the starting point is a relativistic distribution function with the aim to derive rates in the nonrelativistic limit. This is different to what was done in Ref.~\cite{reddy98}. Therefore, the following paragraph summarizes the assumptions and simplifications necessary for the derivation.

A uniform system of only neutrons and protons with RMF interactions is considered, as specified above (momentum-independent interactions, no scalar isovector meson). The approximated single-particle energies of Eq.~(\ref{e_isim}) are used, which employ the (Dirac) effective masses in nonrelativistic kinematics and the potentials of Eq.~(\ref{defui}) and which are valid in the nonrelativistic case $k_i \ll m_i^*$. Then it is straightforward to repeat the calculation of the charged-current rates of Ref.~\cite{bruenn85} within the so-called elastic approximation, where instead of total momentum conservation only the momentum of the nucleons is conserved, ${\bf k}_n = {\bf k}_p$. Because different effective masses of neutrons and protons are considered here, for the derivation one has to assume instead that ${\bf k}_n / \sqrt{m_n^*} = {\bf k}_p /\sqrt{m_p^*}$. For example, for the absorption of a neutrino with energy $\omega$ on a neutron one then obtains:
\begin{eqnarray}
 1/\lambda(\omega)=&& \frac{G^2}{\pi}\eta_{np}(g_V^2+3g_A^2)[1-f_e(\omega+Q')] \nonumber \\ 
&&\times (\omega+Q')^2\left[1-\frac{m_e^2}{(\omega + Q')^2}\right]^{1/2} \nonumber \\
&&\times \theta(\omega-m_e+Q') \; . \label{lambda}
\end{eqnarray}

$Q'$ is the energy release coming from the difference of the single-particle energies of the incoming neutron and the outgoing proton, within the aforementioned approximations (compare with Eq.~(\ref{e_isim})),
\begin{eqnarray}
 E_n-E_p \simeq Q'&=&  m_n-m_p + \Delta U \; ,
\end{eqnarray},
respectively
\begin{eqnarray}
Q' &= &Q+ \Delta U \; .
\end{eqnarray}
It shows that the nucleon potential difference $\Delta U$ leads to a shift in the energy spectrum of the neutrinos. The threshold of $\omega = m_e - Q'$, incorporated in Eq.~(\ref{lambda}) through the $\theta$-function ($\theta(x)=0, x<0; \theta(x)=1, x\geq 0$), is only relevant for proton-rich matter at high densities, namely, if $\Delta U < m_e+m_p -m_n<0$. 

$\eta_{np}$, which originates from the phase-space integrals of the nucleons, is also influenced by the mean-field potentials,
\begin{eqnarray}
 \eta_{np}=(n_n - n_p)/\left(1-\exp[(\mu^0_p-\mu^0_n+ \Delta U)/T]\right) \; , \;\;\;\; \label{eta}
\end{eqnarray}
where
\begin{equation}
 \mu_i^0 = \mu_i-m_i \; ;
\end{equation}
i.e., $\mu_i^0$ is the chemical potential relative to the rest mass.\footnote{Note that one only obtains expression (\ref{eta}) for $\eta_{np}$ if the nonrelativistic Fermi-Dirac integrals give approximately the same number densities as in relativistic kinematics.}
Eq.~(\ref{eta}) can also be written in the following form:
\begin{eqnarray}
   \eta_{np}=(n_n - n_p)/\left(1-\exp[(\nu_p^0 - \nu_n^0)/T]\right) \; , \label{eta2}
\end{eqnarray}
with 
\begin{eqnarray}
 \nu_i^0 = \nu_i-m_i \; .    
\end{eqnarray}
For neutron-rich matter, where $n_n>n_p$, one has $\Delta U > 0$, and also $\nu_n^0>\nu_p^0$. Therefore, one has $\eta_{np}(\Delta U)>\eta_{np}(\Delta U = 0)$, i.e., the overall factor $\eta_{np}$ appearing in the absorptivity, which is independent of the neutrino spectra, is increased by the mean-field potentials. The quantities appearing in Eqs.~(\ref{lambda}) and (\ref{eta}) depend only on $\omega$, $T$, $n_i$, $\mu_i^0$, and $\Delta U$. Thus, one can write
\begin{equation}
 1/\lambda=1/\lambda(\omega, T, \boldsymbol n, \boldsymbol {\mu^0}, \Delta U) \; . \label{deplambda}
\end{equation}

Similarly, one obtains for the emissivity of a neutrino from an electron capture on a proton,
\begin{eqnarray}
 j(\omega)=&& \frac{G^2}{\pi}\eta_{pn}(g_V^2+3g_A^2)f_e(\omega+Q') \nonumber \\ 
&&\times (\omega+Q')^2\left[1-\frac{m_e^2}{(\omega + Q')^2}\right]^{1/2} \nonumber \\
&&\times \theta(\omega-m_e+Q') \; , 
\end{eqnarray}
with
\begin{eqnarray}
 \eta_{pn}&=&(n_n - n_p)/\left(\exp[(\mu^0_n-\mu^0_p- \Delta U)/T]-1\right) \\
 &=& (n_n - n_p)/\left(\exp[(\nu^0_n-\nu^0_p)/T]-1\right) \; .
\end{eqnarray}
The absorptivity for antineutrinos on protons is given by
\begin{eqnarray}
 1/\bar \lambda(\omega)=&& \frac{G^2}{\pi}\eta_{pn}(g_V^2+3g_A^2)[1-f_{\bar e}(\omega-Q')] \nonumber \\ 
&&\times (\omega-Q')^2\left[1-\frac{m_e^2}{(\omega - Q')^2}\right]^{1/2} \nonumber \\
&&\times \theta(\omega-m_e-Q') \; ,
\end{eqnarray}
and the rate for the corresponding emission process is given by
\begin{eqnarray}
 \bar j(\omega)=&& \frac{G^2}{\pi}\eta_{np}(g_V^2+3g_A^2)f_{\bar e}(\omega-Q') \nonumber \\ 
&&\times (\omega-Q')^2\left[1-\frac{m_e^2}{(\omega - Q')^2}\right]^{1/2} \nonumber \\
&&\times \theta(\omega-m_e-Q') \; . 
\end{eqnarray}

\section{Supernova matter}
\label{sec_snmatter}
In SN matter, not only does one have a uniform gas of interacting nucleons, but there is also an important contribution from nuclei. 
This is not only true for the collapse phase and in the accreted matter, where the composition is dominated by heavy nuclei. It was shown in several works \cite{oconnor07, arcones08,sumiyoshi08,hempel11,furusawa2013,fischer14} that light nuclei appear with significant abundances in the envelopes of newly born PNSs. Consequently, the results and derivations presented in the previous section have to be extended to take into account the formation of nuclei. Obviously, in general this is a very complex problem. Here the discussion is restricted to the simplified case that the system can be divided into a uniform nucleon component on the one hand and nuclei on the other, and that separate rate expressions can be applied for the two components. In other words, that the neutrino response is the linear sum of the different contributions. This is actually the standard treatment followed in current CCSN simulations. It is beyond the scope of the present study to provide a more fundamental solution of the problem, e.g., by calculating the neutrino response for the nonuniform and possibly correlated system as a whole. It should be emphasized that neutrino interaction rates with nuclei are not considered in this section. It is only investigated how the presence of nuclei modifies the charged-current neutrino interactions with unbound nucleons. Neutrino interactions with nuclei, where especially electron captures are important, can be found, for example, in Refs.~\cite{langanke03,Juodagalvis:2010,niu2011}.

The EOS model of Ref.~\cite{hempel10}, abbreviated HS in the following, is based on the same underlying, simplifying assumption that is used here for the neutrino interaction rates: Nucleons and nuclei are spatially separated. Consequently, for this model one can achieve a consistent description of the thermodynamic properties and the charged-current neutrino interactions with unbound nucleons. Below it is discussed how the presence of nuclei changes the self-energies, potentials, and elastic charge-current rates of unbound nucleons in this model. For other SN EOSs, the nucleon distributions cannot be reconstructed unambiguously from the published data, which is necessary to derive the local self-energies. Therefore, the discussion is restricted on the EOSs based on the model of Ref.~\cite{hempel10}. Nevertheless, the derivations presented here could serve as a guideline for approximations for other EOSs. 

Obviously, the results and the conclusions which will be drawn are only valid for this particular EOS model. Despite the fact that its nuclear-matter properties are in good agreement with many experimental, theoretical, and astrophysical constraints \cite{fischer14}, and that it is consistent with experimental results for cluster formation in heavy-ion collisions \cite{hempel15}, there are still many aspects of the SN EOS that are rather uncertain. For example, in Ref.~\cite{buyuk13}, the HS model was compared with the EOS models of Furusawa \textit{et al.}~\cite{furusawa2011} and with the statistical multifragmentation model for supernova matter from Botvina and Mishustin \cite{botvina08}. It was demonstrated that the different models show significant differences in their predictions for the abundances of nucleons and light and heavy nuclei. In consequence, already simple thermodynamic quantities and the composition could be rather different using other SN EOS models than the HS model, like the ones mentioned before, or the ones of Refs.~\cite{shen98,lattimer91,shen2011a,shen2011b,typel09,voskresenskaya12,roepke14}. A comparison of the effects of different EOS is, however, beyond the scope of the present investigation.

\subsection{Total self-energies}
\label{sec_totse}
In this section it is first how the presence of nuclei modifies the self-energies of unbound nucleons and quantities that are needed later are defined. In the HS EOS, the total baryon number density $n_B$ is given by
\begin{equation} 
 n_B = n_n + n_p + \sum_k A_k n_k \; ,
\end{equation}
where the sum over $k$ denotes all considered nuclei; i.e., one has $A_k>1$, and $n_k$ is the corresponding number density of nucleus $k$. In HS it is assumed that unbound nucleons occupy only the space which is not filled by nuclei, whereas a volume of $V_k=A_k/n_B^0$ is attributed to each nucleus, with $n_B^0$ being the saturation density of the chosen RMF interactions. Thus, the local number density of the unbound nucleons, i.e., the number of unbound nucleons per free volume, is given by
\begin{eqnarray}
n_n'&=& n_n/\xi \; ,\\
n_p'&=& n_p/\xi \; ,
\end{eqnarray}
with the filling factor $\xi$,
\begin{equation}
 \xi=1-\sum _k V_k n_k =1-\sum _k A_k n_k/n_B^0 \; .
\end{equation}
The excluded volume prescription of the HS model ensures that $0\leq \xi \leq 1$; see Ref.~\cite{hempel10}. Note that in tabulated EOS typically only the global nucleon densities $n_i$ are provided via the mass fractions $X_i=n_i/n_B$, but not the local nucleon densities $n_i'$.

The effective interactions between nuclei and unbound nucleons lead to contributions to the total chemical potentials of unbound neutrons and protons in addition to the RMF interactions. They can be expressed as \cite{hempel10}
\begin{eqnarray}
 \mu_i^{\rm tot}(T,n_B,Y_e) = \mu_i(T,n_n',n_p') + W_i(T,n_B,Y_e) \; . \label{muitot}
\end{eqnarray}
$Y_e$ is the electron fraction which is equal to the total proton fraction $Y_p^{\rm tot}= \frac1{n_B}(n_p + \sum_k Z_k n_k)$, with $Z_k$ denoting the charge number of each nucleus, to obtain charge neutrality. The $\mu_i^{\rm tot}$ are the total chemical potentials, which obey the standard thermodynamic relations for chemical potentials and which are usually provided in tabular EOS. $\mu_i$ are the chemical potentials of the unbound nucleons in the RMF model, as introduced in the previous section and which only depend on temperature and the local number densities of unbound nucleons. In the HS model, $W_i$ contains only contributions from Coulomb and excluded volume interactions.

The total vector self-energies of the unbound nucleons are
\begin{eqnarray}
 \Sigma_V^{i, {\rm tot}}(T,n_B,Y_e) &=& \Sigma_{VR}^{i}(T,n_n',n_p')+ W_i(T,n_B,Y_e) \;\;\;\;\;  \label{vitot}
\end{eqnarray}
and the local Fermi-Dirac distribution functions, i.e., for the unbound nucleons in the free volume, are now given by
\begin{equation}
 f_i = \frac1{1+\exp[(E_i^{\rm tot} -\mu_i^{\rm tot})/T]} \; , \label{disttot}
\end{equation}
with 
\begin{eqnarray}
 E_i^{\rm tot} &=& E_i^{\rm kin}+\Sigma_V^{i, {\rm tot}} = E_i + W_i(T,n_B,Y_e) \; , \label{eitot}
\end{eqnarray}
where Eqs.~(\ref{ei}) and (\ref{vitot}) were used in the last equality. Obviously, the distribution function can also be written as
\begin{equation}
 f_i = \frac1{1+\exp[(E_i -\mu_i)/T]} \; , 
\end{equation}
because the terms $W_i$ from Eqs.~(\ref{muitot}) and (\ref{eitot}) cancel each other. The momentum integration would lead again to the local nucleon number densities $n_n'$ and $n_p'$ of the RMF model, which is an important consistency relation. 

Vice versa, if one wants to calculate $\Sigma_V^{i, {\rm tot}}$, e.g., by using
 \begin{eqnarray}
 \Sigma_V^{i, {\rm tot}}(T,n_B,Y_e) &=& \mu_i^{\rm tot}(T,n_B,Y_e)-\nu_i(T,n_i',\Sigma_S), \; \; \; \; \; \label{zvtot}
\end{eqnarray}
which follows from the previous relations, one has to consider the local nucleon number densities $n_i'$ to calculate $\nu_i$.
In analogy to $\Delta U$, one introduces $\Delta U^{\rm tot}$ (compare with Eqs.~(\ref{defui}) and (\ref{defdu})): 
\begin{eqnarray}
 \Delta U^{\rm tot} &=& U_n^{\rm tot} - U_p^{\rm tot} \label{dutot} \; , \\ 
 U_i^{\rm tot} &=& U_i + W_i \; , \label{uitot}
\end{eqnarray}
giving
\begin{eqnarray}
 \Delta U^{\rm tot} = \Delta \Sigma_V^{\rm tot} = \Sigma_V^{n, {\rm tot}} - \Sigma_V^{p, {\rm tot}} \; .
\end{eqnarray}

\subsection{Results}
\label{sec_rates_tot_results}
\begin{figure*}
\includegraphics[width=1.7\columnwidth]{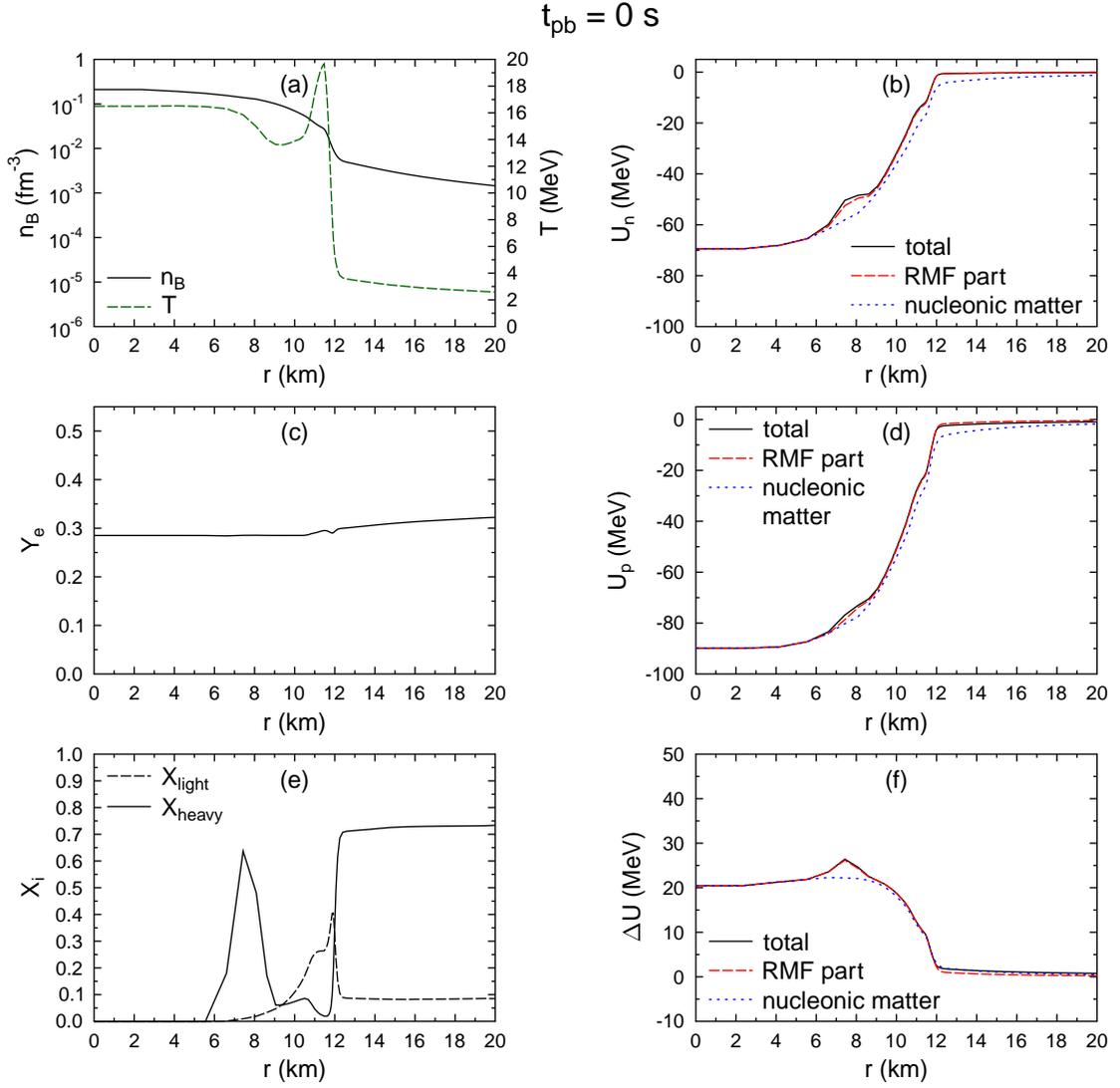}
\caption{(Color online) Left panels, from top to bottom: (a) baryon number density and temperature, (b) electron fraction and (e) mass fractions of light and heavy nuclei as a function of the radius, from a CCSN simulation at bounce using the HS(DD2) EOS \cite{hempel10,fischer14}. Right panels from top to bottom: (nonrelativistic) nucleon potentials of unbound neutrons (b) and protons (d) , and their difference (f). In all of the three right panels the total value (black solid lines) and the RMF contribution (red dashed lines) are shown separately. For comparison, the same quantities are also calculated for the same thermodynamic conditions with the DD2 EOS \cite{typel09}, but employing only nucleons as degrees of freedom (blue dotted lines).}
\label{fig:du_contribs_0}
\end{figure*}
\begin{figure*}
\includegraphics[width=1.7\columnwidth]{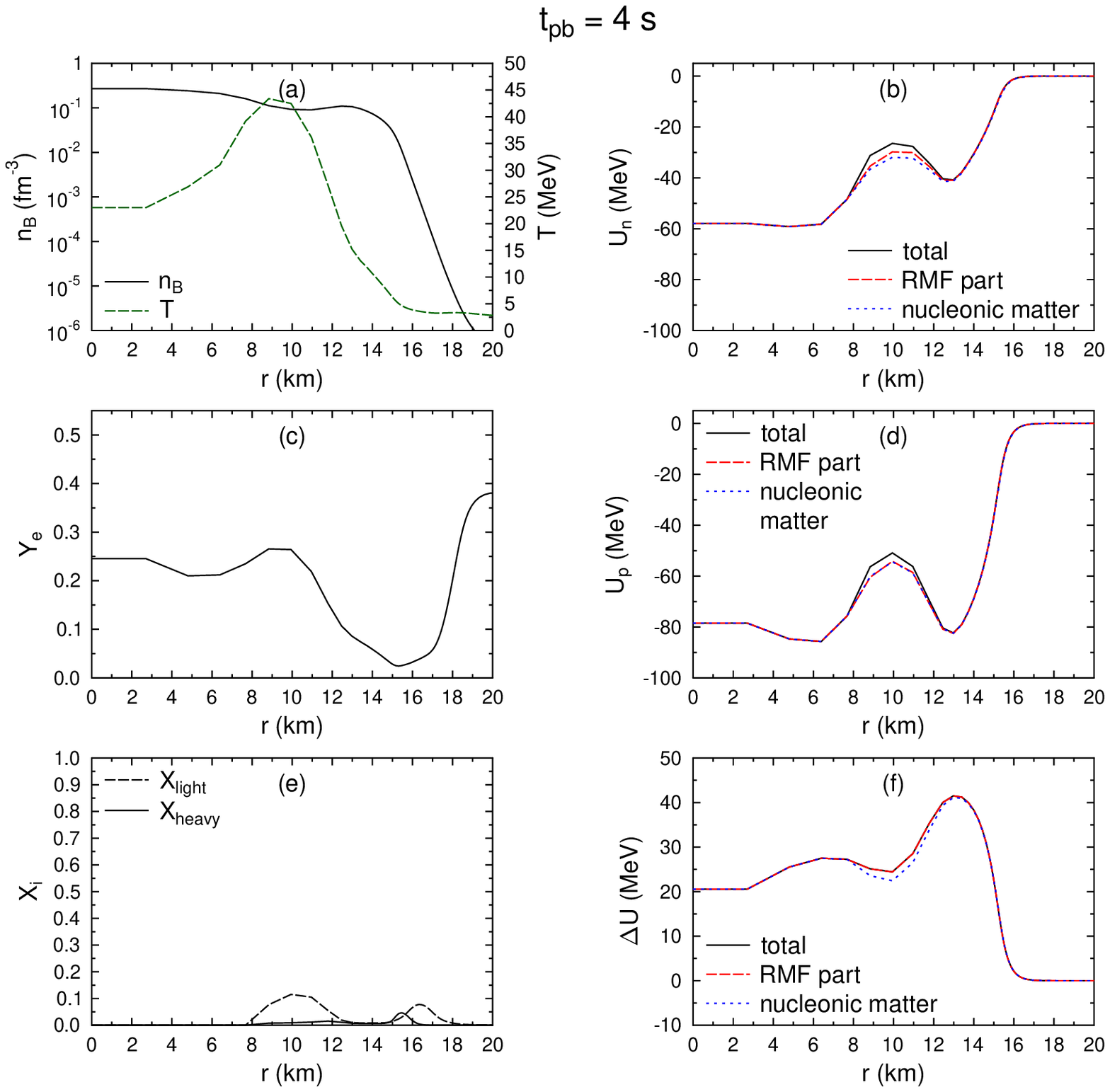}
\caption{(Color online) Same as Fig.~\ref{fig:du_contribs_0}, but at a time of 4~s postbounce.}
\label{fig:du_contribs_4}
\end{figure*}

This section discusses the different contributions to the total self-energies of the unbound nucleons for typical conditions in a CCSN explosion. To do so, the data from the CCSN simulations of Ref.~\cite{perego15} are used. In these simulations artificial explosions are triggered in spherical symmetry to be able to follow the subsequent cooling of the PNS. The simulations are based on the HS(DD2) EOS \cite{hempel10,fischer14}. The left panels of Fig.~\ref{fig:du_contribs_0} show selected thermodynamic properties at core bounce. The right panels show the nonrelativistic nucleon potentials of unbound neutrons and protons and their difference corresponding to this state. The black solid curves are for the total quantities defined by Eqs.~(\ref{uitot}) and (\ref{dutot}). The red dashed curves show only the RMF parts of $U_i$ and $\Delta U$. The blue dotted curves are be explained below.

At this state, the shock is approximately located at a radius of 12~km corresponding to an enclosed baryon mass of 0.6~M$_{\odot}$. In front of the shock, matter consists mostly of heavy nuclei with a minor contribution of light nuclei; see Fig.~\ref{fig:du_contribs_0} (e). Here and in the following, $X_{\rm light}$ is given by the sum of the mass fractions of $\alpha$'s, deuterons, tritons, and helions. $X_{\rm heavy}$ contains all other nuclei. The contribution of light nuclei found here corresponds to mostly $\alpha$ particles. Inside the shock, matter consists mostly of unbound nucleons, besides around 7.5~km, where at densities of $\sim 0.5 n_B^0$ another contribution of heavy nuclei is observed, which is related to the transition to uniform nuclear matter. 

In front of the shock, the unbound nucleon densities are so low, that interactions are almost negligible. The potentials of unbound nucleons are basically zero. For nucleons which are bound in nuclei, of course the potentials keep their typical finite values, but this is not the subject here and it is not shown in the figure. Inside the shock high densities and high mass fractions of unbound nucleons are observed, so that their potentials obtain high values in the range from $-90$ to 0~MeV. One sees that the contribution of nuclei to the unbound nucleon self-energies (i.e., the difference between the black solid and red dashed curves) are generally very small, and only visible around 
the peak of $X_{\rm heavy}$ at 7.5~km. Note that they act repulsively on the unbound nucleons; i.e., they increase their potential. However, this contribution has no visible effect on $\Delta U^{\rm tot}$ because it acts similarly on neutrons and protons.

The blue dotted curves in Fig.~\ref{fig:du_contribs_0} show the same quantities but calculated with the DD2 EOS \cite{typel09} consisting of only nucleons (i.e., without nuclei), for the same density, temperature, and electron fraction profiles. Overall, they lead to a similar qualitative behavior compared to the full calculations including nuclei. For $7< r< 20$~km, they are generally below the total values, because the nucleon densities are higher, owing to the neglect of nuclei. Because of these rather small differences, one can apply the conclusions from Sec.~\ref{sec_exp} also here. Because $Y_e$ inside the shock is roughly constant and the effect of nuclei is small, the behavior of $\Delta U^{\rm tot}$ is approximately set by the behavior of $E^{\rm int, 0}_{\rm sym}$; see Eq.~(\ref{du_approx}). By comparing with Fig.~\ref{fig:eintsym}, one finds the maximum observed for $\Delta U^{\rm tot}$ in Fig.~\ref{fig:du_contribs_0} at similar densities as the maximum of $E^{\rm int, 0}_{\rm sym}$. The only significant difference between $\Delta U^{\rm tot}$ and the 
nucleonic matter case is the additional 
bump on top of this maximum. The reason for the difference is that the appearance of heavy nuclei leads to an increase of the asymmetry of unbound neutrons and protons and thereby to an increase of $\Delta U^{\rm tot}$.

At such an early stage of a SN, the self-energies and potential difference of unbound nucleons have a negligible effect on neutrino quantities and the SN dynamics, because the neutrino spheres are still at very low densities, where nucleon interactions are very weak. This changes in the later evolution, when the neutrino spheres move to higher densities. Figure~\ref{fig:du_contribs_4} shows the same quantities as in Fig.~\ref{fig:du_contribs_0}, but for a stage of 4 s postbounce. To reach this stage, a parameterized explosion was triggered by using the PUSH method as described in Ref.~\cite{perego15}. The details are not important here; for the present purposes it is only important to have a cooling PNS with realistic density, temperature, and electron fraction profiles. 

In Fig.~\ref{fig:du_contribs_4}, the electron fraction shows a local maximum around 9~km. This is related to the high temperatures found here, which lift the electron degeneracy. If one compares the central $Y_e$ with the one at bounce (Fig.~\ref{fig:du_contribs_0}), one can see that it has changed only little, which means that there are still trapped neutrinos and that the PNS is still deleptonizing. Owing to the high temperatures in the range from 5 to 50~MeV, heavy nuclei are not found in the core of the PNS, and light nuclei also only with mass fractions below 0.1. The local maximum in $Y_e$ leads to a local minimum of $\Delta U^{\rm tot}$. The highest potential differences are found around 13~km, at densities around 0.1~fm$^{-3}$ and temperatures of 15~MeV. The neutrino spheres at 4~s postbounce are still located at lower densities (cf.~Ref.~\cite{fischer10a}), but also there high values of $\Delta U^{\rm tot}$ can be expected. The main conclusions from Fig.~\ref{fig:du_contribs_0} remain also valid here: In the HS EOS, nuclei have only a small effect on the potentials of unbound nucleons and these are rather similar to the potentials of a purely nucleonic EOS.

\subsection{Alternative definitions of $\Delta U^{\rm tot}$}
\begin{figure}
\includegraphics[width=0.85\columnwidth]{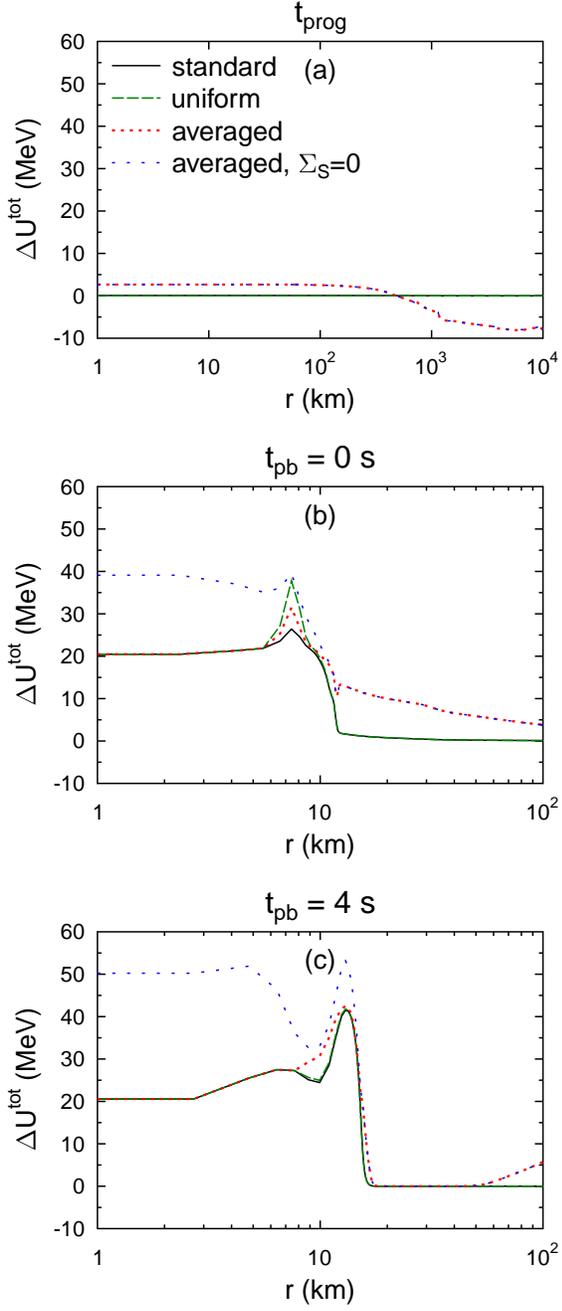}
\caption{(Color online) Different definitions of the potential difference $\Delta U^{\rm tot}$ for a CCSN simulation at the progenitor stage (a), at bounce (b), and at 4~s postbounce (c). For details, see text.}
\label{fig:du_prog}
\end{figure}
In Ref.~\cite{horowitz12}, charged-current neutrino interactions were calculated for the second-order virial EOS which includes deuteron bound-state contributions. It was concluded that the potential difference of the virial EOS is larger compared to standard RMF models. In this study a different definition of the nucleon potentials was used. In the present notation, it would correspond to:
\begin{equation}
 U_i^{\rm tot, av}=\mu_i^{\rm tot} - \mu_i^{\rm free}(T,n_i^{\rm tot}) \; , \label{eq_uihoro}
\end{equation}
whereas $n_i^{\rm tot}$ is the total density of neutrons or protons and given, respectively, by $(1-Y_p^{\rm tot}) n_B$ or $Y_p^{\rm tot} n_B$, and $\mu_i^{\rm free}$ is the chemical potential of a noninteracting Fermi-Dirac gas (i.e., with $m_i^*=m_i$) of neutrons or protons at this density. In the virial EOS, there is no effective mass, and hence in Ref.~\cite{horowitz12} the vacuum mass is used. However, the crucial difference to the definition presented here is that the \textit{total} density $n_i^{\rm tot}$ is used in the second term (given by the sum of bound and unbound neutrons or protons) and not the density of \textit{unbound} neutrons or protons, $n_i$. Because the potential defined in this way obtains contributions of all neutrons, respectively protons, whether bound in nuclei or not, here it is called ``averaged'' potential.

To illustrate the meaning of the definition above, let us consider the case of a noninteracting ideal gas of nucleons and nuclei in nuclear statistical equilibrium (NSE). In this case one has $\mu_i^{\rm tot}=\mu_i^{\rm free}(T,n_i)$ (because of NSE). Thus, the averaged potential is given by:
\begin{eqnarray}
 U_i^{\rm tot, av}&=&\mu_i^{\rm free}(T,n_i) - \mu_i^{\rm free}(T,n_i^{\rm tot}) \; .
\end{eqnarray}
Because $n_i^{\rm tot} > n_i$, one will have $U_i^{\rm tot, av}<0$. With this definition, even for the case of a noninteracting ideal gas of nuclei and nucleons, the presence of nuclei leads to nonvanishing single particle potentials of nucleons, which are attractive. The binding energy contribution of nuclei is averaged over all nucleons. Actually. this is an obvious consequence of the approach used in Ref.~\cite{horowitz12}, because there the distinction between bound and unbound nucleons is not made. Nevertheless, the charged-current rates used in Ref.~\cite{horowitz12} are based on Fermi-Dirac distributions of nucleons. This is plausible for a system consisting only of neutrons, protons and deuterons as it is used in Ref.~\cite{horowitz12}, but not for a system with contributions from strongly bound nuclei such as the $\alpha$ particle or heavy nuclei, which is considered here. 

Let us compare this with the here present standard definition of the potentials for the same case of a noninteracting system. In the present definition, the chemical potential of a free Fermi-Dirac gas of only the unbound, local nucleon contribution is subtracted; i.e., 
\begin{equation}
 U_i^{\rm tot, std}=\mu_i^{\rm tot} - \mu_i^{\rm free}(T,n_i) \; .
\end{equation}
For the ideal noninteracting case this would give
\begin{equation}
 U_i^{\rm tot, std}=\mu_i^{\rm free}(T,n_i) - \mu_i^{\rm free}(T,n_i) =0 \; .
\end{equation}
For a noninteracting system the nucleon potentials are identical to zero, because bound and unbound states are distinguished, and the potentials refer only to the unbound component.

This leads to the conclusion that the increase of the potential difference observed in Ref.~\cite{horowitz12} is based on a different definition of the nucleon potentials. However, the differences in the definitions are only relevant if the bound-state contribution is significant. Note that Ref.~\cite{rrapaj2014} found for typical conditions that the $n$-$p$-scattering resonance in the continuum is more important than the deuteron bound state, which is not addressed here at all.

After the illustrative comments above, a quantitative comparison of various definitions of the potential difference follows. For completeness, also the role of the effective mass is investigated, and the impact of the usage of global instead of local densities of the unbound nucleons. Figure~\ref{fig:du_prog} shows the following four different cases for the same CCSN simulation presented above.
\begin{itemize}
 \item ``standard'' corresponds to $\Delta U^{\rm tot}$ introduced in Sec.~\ref{sec_totse}; see Eq.~(\ref{dutot}). It is considered as the standard definition of the present investigation. It uses the local densities of unbound nucleons and was shown already above in Figs.~\ref{fig:du_contribs_0} (f) and \ref{fig:du_contribs_4} (f) by the black solid lines. Note that compared to Sec.~\ref{sec_rates_tot_results}, the naming convention of this case has been changed from ``total'' to ``standard,'' because of the different context considered here. 
 \item For ``uniform,'' global  densities $n_n$ and $n_p$ of the unbound nucleons are used instead of the local ones $n_n'$ and $n_p'$ in the second term of Eq.~(\ref{zvtot}); i.e., implicitly a uniform distribution of the unbound nucleons is assumed. 
 \item The case ``averaged, $\Sigma_S=0$'' is guided by Ref.~\cite{horowitz12} and uses the total nucleon densities $n_n^{\rm tot}$ and $n_p^{\rm  tot}$ instead, i.e., the sum of bound and unbound nucleons, and does not consider an effective mass. It corresponds to the definition of Eq.~(\ref{eq_uihoro}).
 \item  To investigate the importance of the effective mass, ``averaged'' shows the results if one still uses the total nucleon densities but includes the scalar self-energies of the HS EOS. 
\end{itemize}
In addition to the state at bounce and 4~s postbounce shown above, Fig.~\ref{fig:du_prog} (a) presents also the results for the stage of the progenitor star. 

In Fig.~\ref{fig:du_prog} (a) one sees that the first two definitions give zero potential differences, because the densities are too low. However, for ``averaged'' one sees that it is positive in the center and negative in the outermost layers. Because the effective masses are close to the vacuum values, they do not have any notable impact. The nonvanishing values of $\Delta U^{\rm tot}$ of cases ``averaged''  are generated from the presence of (mostly heavy) nuclei. At bounce, shown in Fig.~\ref{fig:du_prog} (b), one sees the same effect in front of the shock, i.e., for radii above 12 km, compare with Fig.~\ref{fig:du_contribs_0}. In the region around $r\simeq 7.5$~km, where densities are high \textit{and} nuclei are present with significant abundances, all four definitions give different results. Let us explain these differences. If the nucleons are distributed uniformly, they have lower densities than in ``standard,'' and thus $\nu_i(T,n_i)< \nu_i(T,n_i')$. It results in a small difference of $\nu_n$ and $\nu_p$ and therefore to an increased value of $\Delta U^{\rm tot}$ for ``uniform'' compared to ``standard''. The neglect of the effective masses generally increases $\Delta U^{\rm tot}$. Because the nucleon rest masses are always higher than the effective masses, the kinetic chemical potentials entering the definition of the potentials in the case ``averaged, $\Sigma_S=0$'' are dominated by the rest masses. Again it leads to a smaller difference $\Delta \nu_i$ and therefor to an even higher value of $\Delta U^{\rm tot}$ of ``averaged, $\Sigma_S=0$'' compared to ``averaged.''

In the postexplosion phase, where the NDW is generated, the abundances of nuclei in the relevant density range are lower, as can be seen in Fig.~\ref{fig:du_contribs_4} (e). Consequently, in Fig.~\ref{fig:du_prog} (c), the difference between local and global nucleon densities (i.e., between the cases ``standard'' and ``uniform'') is not so important anymore. However, the averaged nucleon potential difference is still notably larger. Also the effect of the effective masses is enhanced, owing to the higher densities reached.

The conclusion is that ``standard'' and ``uniform'' give similar results for most conditions. Note that for other existing SN EOS, it would not be possible to calculate something equivalent to ``standard'' in an exact way, because the information about the local nucleon distribution functions is typically not provided and cannot be reconstructed completely. However, based on the findings presented here, the potentials corresponding to ``uniform,'' which always can be calculated, can be taken as a first approximation. The case ``averaged'' leads to higher values of the nucleon potentials. Even in the noninteracting regime it can have nonzero values, because it is based on a qualitatively different picture, where bound states of nuclei are not treated separately. As long as one uses charged-current neutrino interaction rates that are based on Fermi-Dirac distribution functions of nucleons, it seems to be more consistent to use the potentials of ``standard'' instead.

The different definitions of the potential differences ``standard'' and ``averaged'' can be related to different definitions of the symmetry energy. It was shown in Sec.~\ref{sec_rates_tot_results} that the results for $\Delta U^{\rm tot}$ for ``standard'' are relatively similar to those of nucleonic matter. Therefore, $\Delta U^{\rm tot}$ is approximately given by the potential part of the symmetry energy of the unbound nucleon component, as discussed in Sec.~\ref{sec_nucleonic}. ``averaged,'' however, contains direct bound-state contributions. It could possibly be related to the symmetry energy of clusterized matter \cite{natowitz10,typel14}, where the binding energies of clusters contribute directly, too.

\subsection{Elastic charged-current rates}
\label{sec_rates_tot}
Next, the weak charged-current rates with the unbound nucleon component are discussed for the HS SN EOS model in the case where also nuclei are present. Let us start with neutrino absorption on unbound neutrons. Within the simplified geometrical picture, which is employed here, the total absorptivity has to be weighted with the filling factor $\xi$,
\begin{eqnarray}
 1/\lambda^{\rm tot} = \xi \, 1/\lambda \; , \label{eq_def_ltot}
\end{eqnarray}
because only the fraction $\xi$ of the total volume is filled with these neutrons. $1/\lambda$ is the absorptivity inside the free volume.

However, inside the free volume, the unbound nucleons still obey Fermi-Dirac statistics, with the only difference that their chemical potentials and self-energies have also a contribution from the interactions with nuclei. Thus, the only thing to do, if one starts from the distribution function (\ref{disttot}) and compares with Eq.~(\ref{fi}), is to replace $\boldsymbol n$, $\boldsymbol{\mu^0}$, and $\Delta U$ in the expression (\ref{deplambda}) for $1/\lambda$ with $\boldsymbol {n'}$, $\boldsymbol {\mu^{\rm tot 0}}$, and $\Delta U^{\rm tot}$, respectively. ${\mu_i^{\rm tot 0}}$ is defined as ${\mu_i^{{\rm tot}}}-m_i$. Thus, owing to the presence of nuclei, Eq.~(\ref{deplambda}) changes to
\begin{equation}
 1/\lambda=1/\lambda(\omega, T, \boldsymbol {n'}, \boldsymbol {\mu^{\rm tot 0}}, \Delta U^{\rm tot}) \; . \label{deplambda_tot}
\end{equation}
Formulated in this way, to calculate $1/\lambda^{\rm tot}$, one still had to know $\xi$ in addition, which appears in Eq.~(\ref{eq_def_ltot}).

However, the expression for the total absorptivity can be simplified further. It is useful to introduce $\eta_{np}^{\rm tot} = \xi \eta_{np}$. Then the full expression for $1/\lambda^{\rm tot}$ can be expressed as
\begin{eqnarray}
 1/\lambda^{\rm tot}(\omega)=&& \frac{G^2}{\pi}\eta^{\rm tot}_{np}(g_V^2+3g_A^2)[1-f_e(\omega+Q'^{\rm tot})] \nonumber \\ 
&&\times (\omega+Q'^{\rm tot})^2\left[1-\frac{m_e^2}{(\omega + Q'^{\rm tot})^2}\right]^{1/2} \nonumber \\
&&\times \theta(\omega-m_e+Q'^{\rm tot}) \; , 
\end{eqnarray}
whereas 
\begin{equation}
 Q'^{\rm tot}=Q + \Delta U^{\rm tot} \; .
\end{equation}
Written explicitly, $\eta_{np}^{\rm tot} = \xi \eta_{np}$ is given as
\begin{eqnarray}
   \eta_{np}^{\rm tot}&=&\xi(n_p' - n_n')/\left(\exp[(\nu_p^0 - \nu_n^0)/T]-1\right) \\
&=& (n_p - n_n)/\left(\exp[(\nu_p^0 - \nu_n^0)/T]-1\right) \; .
\end{eqnarray}
This can also be written as
\begin{eqnarray}
   \eta_{np}^{\rm tot}&=&(n_p - n_n)/\left(\exp[(\mu_p^{\rm tot 0} - \mu_n^{\rm tot 0}-\Delta U^{\rm tot})/T]-1\right) \; , \nonumber \\
\end{eqnarray}
by using Eq.~(\ref{zvtot}). The filling factor $\xi$ does not appear anymore. If one compares with Eq.~(\ref{eta}), one sees that the nucleon chemical potentials and the nucleon potential difference are simply replaced with the corresponding total quantities. In conclusion, the absorptivity can be calculated directly from $\omega$, $T$, $n_i$, $\mu_i^{\rm tot 0}$, and $\Delta U^{\rm tot}$,
 \begin{equation}
 1/\lambda^{\rm tot}=1/\lambda(\omega, T, \boldsymbol {n}, \boldsymbol {\mu^{\rm tot 0}}, \Delta U^{\rm tot}) \;  \label{deplambda_tot2} \; . 
\end{equation} 
The temperature, densities, and total chemical potentials are usually part of EOS tables; thus, the only additional quantity which is needed for the consistent rates as specified here is $\Delta U^{\rm tot}$. 

In the same way, one obtains for the emission rate of neutrinos from electron captures on unbound protons,
\begin{eqnarray}
 j(\omega)^{\rm tot}=&& \frac{G^2}{\pi}\eta_{pn}^{\rm tot}(g_V^2+3g_A^2)f_e(\omega+Q'^{\rm tot}) \nonumber \\ 
&&\times (\omega+Q'^{\rm tot})^2\left[1-\frac{m_e^2}{(\omega + Q'^{\rm tot})^2}\right]^{1/2} \nonumber \\
&&\times \theta(\omega-m_e+Q'^{\rm tot}) \; , 
\end{eqnarray}
with
\begin{eqnarray}
 \eta_{pn}^{\rm tot}&=&(n_n - n_p)/\left(\exp[(\mu^{\rm tot 0}_n-\mu^{\rm tot 0}_p- \Delta U^{\rm tot})/T]-1\right) \; . \nonumber \\
\end{eqnarray}

The rate for absorption of antineutrinos on unbound protons is given as
\begin{eqnarray}
 1/\bar\lambda^{\rm tot}(\omega)=&& \frac{G^2}{\pi}\eta_{pn}^{\rm tot}(g_V^2+3g_A^2)[1-f_{\bar e}(\omega-Q'^{\rm tot})] \nonumber \\ 
&&\times (\omega-Q'^{\rm tot})^2\left[1-\frac{m_e^2}{(\omega - Q'^{\rm tot})^2}\right]^{1/2} \nonumber \\
&&\times \theta(\omega-m_e-Q'^{\rm tot}) \; ,
\end{eqnarray}
and the rate for the corresponding emission process,
\begin{eqnarray}
 \bar j(\omega)^{\rm tot}=&& \frac{G^2}{\pi}\eta_{np}^{\rm tot}(g_V^2+3g_A^2)f_{\bar e}(\omega-Q'^{\rm tot}) \nonumber \\ 
&&\times (\omega-Q'^{\rm tot})^2\left[1-\frac{m_e^2}{(\omega - Q'^{\rm tot})^2}\right]^{1/2} \nonumber \\
&&\times \theta(\omega-m_e-Q'^{\rm tot}) \; . 
\end{eqnarray}

Next one can derive detailed balance to be
\begin{eqnarray}
 1/\lambda^{\rm tot}(\omega) &=& \exp{ \left\{ [\omega -(\mu_p^{\rm tot}+\mu_e-\mu_n^{\rm tot})]/T \right\} } j(\omega)^{\rm tot} \; , \nonumber \\
\\
1/\bar\lambda^{\rm tot}(\omega) &=& \exp{ \left\{ [\omega -(\mu_n^{\rm tot}-\mu_p^{\rm tot}-\mu_e)]/T \right\} } \bar j(\omega)^{\rm tot} \; . \nonumber \\
\end{eqnarray}
It shows that the charged-current rates with unbound nucleons drive the system to the correct global weak equilibrium. Emissivity and absorptivity become equal for thermalized neutrinos if $\mu_n^{\rm tot}+\mu_{\nu_e} =\mu_p^{\rm tot}+\mu_e$. Note also that if nuclei are not present, the derived rates are identical to the pure mean-field expressions from Sec.~\ref{sec_rates_rmf}, because in this case $\xi=1$, $U_i=U_i^{\rm tot}$, and $\mu_i=\mu_i^{\rm tot}$.

\section{Summary and Conclusions}
\label{sec_summary}
This article investigates nucleon self-energies in SN matter and provides corresponding basic expressions for charged-current neutrino interaction rates with unbound nucleons. The presented work is essentially motivated by Refs.~\cite{martinez12,roberts12}, where it was shown that the difference of the neutron and proton interaction potentials has an impact on neutrino spectra in the neutrino-driven wind (NDW) phase of CCSNe, which is very important for the related nucleosynthesis. 

In the first part of the article, the contribution of heavy nuclei was neglected and solely nucleonic RMF models were investigated. The used formalism, based on the scalar and vector self-energies, allowed a rather general discussion, with the restrictions that no scalar isovector interactions (in meson-exchange models, the $\delta$ meson) were included and that the interactions were chosen to be momentum independent (beyond the standard dependence via the effective mass). It was shown that the quadratic approximation of the EOS works reasonably well at finite temperature. Deviations are visible at high temperatures and/or densities that originate almost entirely from the kinetic contribution. The use of realistic nucleon masses leads to an important linear term, which otherwise would not be present. Furthermore, it was shown that the interaction part of the second-order coefficient in the expansion, the so-called interaction symmetry energy $E_{\rm sym}^{\rm int}$, is almost temperature independent for the models considered here. This is supported by Ref.~\cite{fedoseew2014}, which showed that the temperature dependence of the nucleon self-energies is negligible by comparing with Dirac-Brueckner calculations.

This is in contrast to the kinetic contribution $F_{\rm sym}^{\rm kin}$, which is very sensitive to temperature. It was derived that the difference of the vector self-energies of neutrons and protons in first order is proportional to the asymmetry $1-2Y_p$ and $E_{\rm sym}^{\rm int}$; see Eq.~(\ref{yeah}). Higher-order terms in $Y_p$ were found to be small or absent.
This equation is an important result of the present investigation
and refines previous purely qualitative statements about the connection between the symmetry energy and the nucleon potentials. In consequence, for the RMF models considered here, $\Delta U$ is almost temperature independent, because of the approximate temperature independence of $E_{\rm sym}^{\rm int}$. 
Models with a high $E_{\rm sym}^{\rm int}$ typically also have a high free symmetry energy $F_{\rm sym}=F_{\rm sym}^{\rm kin}+E_{\rm sym}^{\rm int}$, which, in turn, results in a high $Y_e$ (i.e., closer to 0.5) in $\beta$-equilibrated matter. In principle, this could lead to a compensation effect in $\Delta U$. However, it was found that even for NS matter, i.e., for $T=0$ and $\beta$ equilibrium without neutrinos, this compensation effect is not dominating, i.e., the shape of $\Delta U$ still resembles the one of $E_{\rm sym}^{\rm int}$. 

Different RMF models were compared with the experimental constraints for the (zero temperature) symmetry energy of Refs.~\cite{danielewicz2013,lattimer2013}. Strictly speaking, it is clear that these constraints cannot be applied directly on the interaction part of the symmetry energy $E_{\rm sym}^{\rm int}$ alone (which determines $\Delta U$), but only on the total symmetry energy $E_{\rm sym}$. Nevertheless, because the kinetic contribution $E_{\rm sym}^{\rm kin}$ is rather similar for all the considered models at low densities, the experimental results still can be used to constrain the behavior of $E_{\rm sym}^{\rm int}$ and therefore also of $\Delta U$ at low densities. The EOS of LS, and the simple nonlinear RMF models NL3, TM1, and TMA show a large discrepancy from the experimental constraints. This is in line with the conclusions from Ref.~\cite{fischer14} and also with Ref.~\cite{dutra2014}, regarding the simple nonlinear RMF models. The best agreement was found for DD2, FSUgold, SFHo, and SFHx. Note, however, that FSUgold is excluded by astrophysical observations of 
NSs \cite{fischer14}. Compared to TM1, which is also employed in the commonly used EOS of STOS \cite{shen98}, these more modern density functionals give higher values of $\Delta U$ at subsaturation densities. This is the density region which is most relevant for the neutrino spheres during the NDW phase. IUFSU is the only model whose symmetry energy at these densities is too high compared with the experimental constraints. Therefore, its corresponding values of $\Delta U$ can be interpreted as overestimated.

In the second part, the role and effect of nuclei on single-particle properties of the unbound nucleons was investigated. The derivations were restricted to SN EOS, which are based on the HS model \cite{hempel10}. Nevertheless, they can also serve as guidelines for other models. It was shown that in addition to the RMF contributions, also the interactions with nuclei have an effect on the self-energies of unbound nucleons. In the HS model, these are mostly excluded volume interactions, and for certain conditions also Coulomb interactions. However, regarding the potential difference of the unbound nucleons, the former interactions are equal for neutrons and protons and therefore do not contribute. Obviously, this could be different in other EOS models. It was also found that the self-energies of a purely nucleonic RMF model show a qualitatively similar behavior compared to the full calculation {of the self-energies of unbound nucleons} including nuclei. Therefore, one can expect only minor changes regarding the neutrino emissivities and absorptivities with reactions on unbound nucleons compared to a purely nucleonic EOS, as was, e.g., done in
Ref.~\cite{roberts12a}. However, it should also be stressed that the contributions of nuclei to the neutrino interaction rates still could lead to significant changes. This was not addressed in the present study. Furthermore, despite the effect of nuclei on unbound nucleons was shown to be small, the results presented here give a more consistent description between charged-current rates with unbound nucleons and the thermodynamic properties of the EOS. Electronic tables with the self-energies of the unbound nucleons are provided online (see footnote \ref{eospage}) for eight different RMF models.

There are already several works in the literature which investigated the effect of the nucleon potentials on the asymptotic electron fraction in the NDW. Fischer \textit{et al.}\ obtained a minimal $Y_e$ of 0.48 \cite{martinez14} using the HS(DD2) EOS. Roberts \textit{et al.}\ considered the IUFSU and GM3 \cite{glendenning91} interactions and obtained minimal $Y_e$ values of 0.46 and 0.50.\footnote{Note that the two numbers are different compared to Ref.~\cite{roberts12}, owing to a previous computational error which was now corrected (L.\ Roberts, presentation at the MICRA workshop in Trento, 2013).} GM3 was not included in the present investigation. However, GM3 has values of $\Delta U$ lower than those of NL3 and slightly higher values than those of TMA, which is the lowest curve of Fig.~\ref{fig:du}. IUFSU on the contrary, gives the highest values of all models, and DD2 is right in the middle. Thus, one can conclude that these three simulations have already probed the range of $\Delta U$ from RMF models that is consistent with nuclear experiments. Even by taking the highest potential difference of IUFSU, the minimal $Y_e$ obtained is only 0.46, which would not allow a full r-process.

Finally, alternative definitions for the potential difference of nucleons were compared with each other, which relate to different treatments of nuclei. In one case, the global instead of the local densities of unbound nucleons were used. This is interesting, because only the former, but not the latter quantity is typically provided for other existing EOS tables such as the LS or STOS EOS. Obviously, the distinction between local and global nucleon densities is only relevant if nuclei are abundant; otherwise, they are identical. It was found that the total potential difference could be slightly overestimated, if it was calculated from the global nucleon densities, but the differences are not extreme and the overall behavior is reproduced well.

As another case, a definition of the potential difference similar to the one proposed in Ref.~\cite{horowitz12} was considered. In Ref.~\cite{horowitz12} the potential difference was calculated for the second-order virial EOS including the deuteron bound state. The nucleon potentials in this definition have a direct contribution of the bound states via their binding energies. Consequently, for systems that contain strongly bound nuclei it does not lead to vanishing nucleon potentials even at low densities, which is in contrast to the standard definition proposed here. 
In the standard definition, binding energies of nuclei do not contribute to the potentials of unbound nucleons directly. In consequence, the effect of nuclei is weak and the total potential difference is approximately given by the potential part of the nucleonic symmetry energy. In the definition that is similar to the one of Ref.~\cite{horowitz12} the opposite is the case, and the potential difference is more related to the symmetry energy of clusterized matter; see, e.g., \cite{natowitz10,typel14}. 

To arrive at a more conclusive comparison between RMF models and the virial EOS, it would be necessary to further disentangle the effect of unbound, bound, and scattering states. It will also be important to further compare the predictions of RMF models with many-body calculations employing realistic nucleon interactions. Regarding investigations on the mean-field level, it would be interesting to consider the effective mass splitting of neutrons and protons (see, e.g., Ref.~\cite{fedoseew2014}) or also new momentum-dependent interactions, as, e.g., the ones of Ref.~\cite{gaitanos2013}.

It is clear that the underlying picture used in the present approach, that the neutrino response is the linear sum of the contributions of unbound nucleons and nuclei, is too simplified for certain conditions. The emergence of different definitions of the potentials in the literature and the discussion above simply illustrates the complexity of the SN EOS, if one requires to have a unified description of thermodynamic and microscopic quantities from the collapse of the progenitor star until the stage of the cold NS. The change of the degrees of freedom between heavy and light nuclei and nucleons represents a severe complication. It was shown here that the purely nucleonic component (on the mean-field level) is rather well under control and also constrained experimentally at low densities. The theoretical description of the bound and scattering states is a much more complex problem, as can also be seen, e.g., in Refs.~\cite{rrapaj2014,roepke14}. Fortunately, heavy-ion collision experiments can be used to probe the formation of nuclei in SN matter (see, e.g., \cite{natowitz10,qin12,hempel15}) which helps to constrain the theoretical models. 

\subsection*{Acknowledgments}
I thank T.~Fischer, G.~Mart{\'{\i}}nez-Pinedo, M.~Liebend\"orfer, and F.-K.~Thielemann for their helpful comments and discussion of this work. Support from the Swiss National Science Foundation (SNSF) is greatly acknowledged. Partial support comes from ``NewCompStar,'' COST Action MP1304. I am also grateful for participating in the EuroGENESIS collaborative research program of the ESF and the ENSAR/THEXO project.

\appendix
\section*{Appendix: Tables with self-energies and other microscopic quantities}
For the different SN EOS tables discussed in this article---SFHo, SFHx, HS(TMA), HS(TM1), HS(FSUgold), HS(IUFSU), HS(NL3), and HS(DD2)---electronic data tables are provided\footnote{See \texttt{http://phys-merger.physik.unibas.ch/\midtilde hempel/eos.html}\label{eospage}.} containing the following information:

\begin{enumerate}
 
\item baryon number density $n_B$ (fm$^{-3}$);

\item total proton fraction $Y_p^{\rm tot}$ (dimensionless);

\item total vector self-energy of unbound neutrons $\Sigma_V^{\rm n,tot}$ (MeV);

\item total vector self-energy of unbound protons $\Sigma_V^{\rm p,tot}$ (MeV);

\item filling factor of unbound nucleons $\xi$ (dimensionless);

\item effective Dirac mass of unbound neutrons $m_n^*$ (MeV);

\item effective Dirac mass of unbound protons $m_p^*$ (MeV).

\end{enumerate}
In combination with the information provided in the EOS tables (e.g., $X_i$, $\mu_i^{\rm tot}$), it is possible to derive all quantities presented in this article and to calculate the charged-current rates, e.g., using the expressions of Sec.~\ref{sec_rates_tot}.

The data are arranged in the following way: They are grouped in blocks of constant temperature, starting with lowest values. Within each temperature block, the data are grouped according to the proton fraction, again starting with lowest values. For given temperature and proton fraction all baryon number densities are then listed with increasing values. 

\bibliographystyle{apsrev}
\bibliography{literat}

\end{document}